\documentstyle[aps,psfig,doublespace]{revtex}
\begin{document}

\title{{\bf Polarisation Patterns and Vectorial Defects in Type II
Optical Parametric Oscillators}}

\author{M. Santagiustina$^{1}$, E. Hernandez-Garcia$^{2}$, M.
San-Miguel$^{2}$,
A.J. Scroggie$^{3}$, G.-L. Oppo$^{3}$\\
\vskip 0.5cm
$^{1}$ Istituto Nazionale di Fisica della Materia (INFM),\\
Dipartimento di Elettronica e Informatica,\\
Universit\'a di Padova, I-35131, Padova, Italy\\
$^{2}$ Instituto Mediterraneo de Estudios Avanzados (IMEDEA), \\
CSIC-Universitat de les Illes Balears\\
E-07071 Palma de Mallorca, Spain\\
$^{3}$ Department of Physics and Applied Physics,\\
University of Strathclyde,\\
107 Rottenrow, Glasgow G4 0NG, UK.}

\maketitle
\vskip 1cm

\begin{abstract}
Previous studies of lasers and nonlinear resonators
have revealed that the polarisation degree of freedom allows for the
formation of polarisation patterns and novel localized structures, such
as vectorial defects. Type II optical parametric oscillators are characterised
by the fact that the down-converted beams are emitted in orthogonal
polarisations. In this paper we show the results of the study
of pattern and defect formation and dynamics in a Type II degenerate optical
parametric oscillator for which the pump field is not resonated in the
cavity. We find that traveling waves are the predominant solutions and
that the defects are vectorial dislocations which appear at the boundaries
of the regions where traveling waves of different phase or wave-vector
orientation are formed. A dislocation is defined by two topological charges, one
associated with the phase and another with the wave-vector orientation. We also
show how to stabilize a single defect in a realistic experimental
situation. The effects of phase mismatch of nonlinear interaction are
finally considered.
\end{abstract}

\section{Introduction}

Most studies of pattern \cite{chaos1,chaos2,pinos} and localized structure
\cite{Tlidi94,Saffman94,Firth96,Rosanov96,Schreiber96,Peschel97,Staliunas98,Spinelli98,Weiss99,Oppo99,LeBerre99,Ackemann00,Ramazza,oppo98}
formation in optical systems consider light with a fixed linear
polarisation. However, the vectorial degree of freedom of light,
associated with a space- and time-dependent polarisation, leads
to a very interesting phenomenology, as previously shown for
Kerr cavities \cite{gedd94,Hoyuelos98,rafa99}, sodium vapors \cite{auma97},
lasers \cite{maxi95,jose97,HernandezGarcia99} and intra-cavity second harmonic
generation \cite{Peschel98}. This degree of freedom is also relevant
from the point of view of information encoding and processing, because
it yields a tool for the control of these structures.
Optical vortices have been predicted to occur in optical systems such as lasers
\cite{coul89} and they have been experimentally observed in lasers \cite{coates94}
and photo-refractive resonators \cite{Weiss99,arech91,stal97}. More recent
theoretical studies revealed the existence of vectorial defects in optical
systems \cite{HernandezGarcia99}.

Among optical systems, optical parametric oscillators (OPO's) have been the
object of intense theoretical study in recent years, which has revealed the
existence of patterns \cite{oppo94,oppo94a,stal95,long96,long96a,long96b,long96c,valc96,valc97,pre,taki98,saff00}
and localized structures \cite{Staliunas98,Oppo99,LeBerre99,oppo98,tril97,long97,tril98,ol};
few examples of defect dynamics have also been published \cite{oppo98,valc97,josab}.
Finally, experimental results on transverse pattern formation
have been recently presented \cite{fabre99,ducci01}. An OPO consists of a
ring (or Fabry-Perot) resonator containing a nonlinear quadratic
medium, which performs a parametric down-conversion of an injected
laser beam (optical pump) at frequency $\omega_p$. Two new fields are
generated in the crystal, the signal and the idler, at frequencies
$\omega_s, \omega_i$ respectively, such that $\omega_p=\omega_s+\omega_i$, the
last relation indicating the conservation of photon energy. The nonlinear
interaction also requires photon momentum to be conserved, i.e. the phase-matching
condition $k_p-k_s-k_i = 0$, which is usually obtained by exploiting polarized
beams and crystal birefringence. This effect often implies that a
transverse walk-off between the different polarisation components may
occur. The consequences of walk-off have also been considered and
convective instabilities and noise-sustained patterns found
\cite{pre,taki98,ol,oe,taki00}. Most of these theoretical studies have been performed in
the context of a scalar approximation, which is valid for the so called
Type I OPO's; in the Type I interaction the signal and idler
fields have the same polarisation, orthogonal to the pump beam.
However, there are not many studies involving Type II OPO's where the
polarisation degree of freedom is taken into account. In refs.
\cite{oppo94a,long96a,long96b,valc97} the model equations can be valid
for an OPO that is non degenerate either in polarisation or in frequency;
however the intrinsic physical difference between the two
cases was not made evident. In refs. \cite{josab} and \cite{ward01} the
polarisation non-degenerate case and the polarisation and frequency
non-degenerate case have been considered and the role of the walk-off explored.
Finally the case of polarisation coupling, due to cavity birefringence and/or
dichroism has been analysed for the case of an OPO, showing Bloch walls formation
\cite{ol00}; these structures were later found also for second harmonic generation
\cite{mich01}.

In this paper we will focus our attention on the polarisation pattern
and vectorial defect formation in a Type II, frequency degenerate OPO.
In particular we will consider the case in which the pump is
not resonated in the cavity and can be eliminated from the dynamics by
means of a multiple scale approximation. This simplification has also
been made in order to take into account the effects of phase mismatch,
which has been only partly studied in connection with Type I OPO's
\cite{long96}. In fact, the conservation of the photon momentum need not
to be exact for down conversion to occur, the conversion efficiency being
lower if $k_p-k_s-k_i \neq 0$.

The paper is organized as follows. In section \ref{equat} we
present the equations which model the Type II OPO and some exact,
stable solutions, which are combinations of traveling waves (TW's). Due
to the vectorial nature of the field these solutions represent spatial
polarisation patterns. We also show that these TW's are spontaneously formed
starting from a random perturbation of the trivial steady-state, that is
represented by the field below the threshold of signal and idler generation.
In section \ref{defe} we show that, on a background made of TW's, vectorial defects
are spontaneously generated. The defects are classified as vectorial dislocations by
comparing the quantities characteristic of a dislocation of a complex field with
those of the defects obtained in the dynamical system. The dynamics of defects
is also studied in this section. The stabilization of a
single defect is then presented in section \ref{isol}.  The discussion
of the effects due to the presence of the phase mismatch is given in
section \ref{frem} and the conclusions presented in section
\ref{concl}. In Appendix I we derive the basic equations in which our
study is based, and in Appendix II we derive the corresponding amplitude
equation (a Swift-Hohenberg equation) for the instability leading to
frequency-converted light generation in our model.

\section{Equations and background solutions}
\label{equat}

We considered a model for a Type II OPO, for which the pump is not
resonated in the cavity and in which some mismatch among the
interacting waves is allowed. Previous studies of pattern formation,
except ref. \cite{long96}, neglected mismatch; however real devices are
likely to present a residual mismatch due to the selection mechanism of the
signal and idler frequencies (see \cite{ecka91} for details).  Since
the Type II configuration means that the signal and the idler are
orthogonally polarized fields we will use the notation $A_x, A_y$ to
denote them. The starting point is the slowly varying envelope
approximation propagation equations for a Type II interaction which
include the mismatch (see for example \cite{zhan95}). By following the
procedure outlined in Appendix I, we obtain the evolution equations for
the cavity fields:

\begin{eqnarray}
\label{master1}
\partial_t A_x &=& \gamma_x [ -(1+i \Delta) A_x
+ \mu A_y^* + \sigma |A_y|^2 A_x] + i a_x \nabla^2 A_x  \\
\label{master2}
\partial_t A_y &=& \gamma_y [ -(1+i \Delta) A_y
+ \mu A_x^* + \sigma |A_x|^2 A_y] + i a_y \nabla^2 A_y
\end{eqnarray}

where $\gamma_x,\gamma_y$ are the decay rates of the signal and idler
in the cavity, $\Delta$ the cavity detuning, $a_x, a_y$ the diffraction
coefficients. Note that the detunings of the two fields are made equal to a
common value $\Delta$ \cite{DSGF} by choosing the temporal reference frame
such that stationary homogeneous states do not have free rotating phases.
The other terms which appear in these coupled equations are respectively:

a) a linear parametric coupling with coefficient
\begin{equation}
\label{mu}
\mu=E_0 exp(\frac{i \Delta k L}{2}) \frac{sin (\Delta k L/2)}{\Delta k L/2}
\end{equation}
where $E_0$ is the injected pump field, $\Delta k$ the photon momentum
mismatch and $L$ the crystal length \cite{note};

b) a cubic nonlinear coupling with coefficient
\begin{equation}
\label{sigma}
\sigma=\frac{2}{\Delta k^2 L^2} [ (cos(\Delta k L)-1) +
i (sin(\Delta k L) -\Delta k L)]
\end{equation}

At perfect phase-matching $\mu=E_0, \sigma=-1$ ; otherwise the coefficients are
complex numbers; their real, imaginary parts are shown in fig. \ref{reimcoef} for
$E_0=1.5$.  Similar coefficients were also found for a Type I, singly resonant
frequency degenerate OPO, by Longhi in ref. \cite{long96}, with the same type of
analysis. Note that the lasing of an OPO requires $|\Delta k| < \pi/L$ because the
parametric gain outside this range is too small and the device tends to "jump" to
another pair of signal-idler frequencies which satisfy this condition
\cite{ecka91}.
Finally note that all coefficients can become dimensionless as soon as appropriate
time, space and field amplitude scaling are performed.

Equations (\ref{master1},\ref{master2}) have exact traveling wave solutions of
the same type as those found in ref. \cite{long96b}:

\begin{equation}
\label{tw}
\left( \begin{array}{c} A_x \\
A_y^* \end{array} \right)=\sqrt{C} e^{i(\omega t+{\bf k} \cdot {\bf r})}
\left( \begin{array}{c} e^{i \phi_0} \\
e^{-i (\alpha-\phi_0)} \end{array} \right)
\end{equation}

where $\phi_0$ is an arbitrary phase, while the intensity $C$ and the phase
difference $\alpha$ between the signal and the idler are given by

\begin{eqnarray}
\label{det-c}
C=\frac{1}{|\sigma|^2} \left[ \sigma'+\sigma'' \tilde{\Delta} \pm \sqrt{|\mu|^2
|\sigma|^2 - (\sigma'' - \sigma' \tilde{\Delta})^2} \right] \\
\label{det-psi}
cos \alpha=\frac{1}{|\mu|^2} [\mu' (1-\sigma' C) + \mu'' (\tilde{\Delta} -
\sigma'' C)]
\end{eqnarray}

where $\sigma'=Re(\sigma), \sigma''=Im(\sigma)$,
$\mu'=Re(\mu),\mu''=Im(\mu)$ and $\tilde{\Delta}= \Delta + |{\bf k}|^2
(\gamma_x a_x+\gamma_y a_y)/(\gamma_x+\gamma_y)$ is an effective detuning
parameter. The TW intensity $C$ and the phase difference $\alpha$ are shown in
fig. \ref{c-psi} for $E_0=1.5$.

The frequency shift $\omega$ is given by:

\begin{equation}
\omega=\frac{\gamma_x \gamma_y}{\gamma_x+\gamma_y}
[(a_y-a_x)|{\bf k}|^2]
\end{equation}

This quantity can be zero if the diffraction coefficients are equal
for the signal and the idler. Hereafter we will consider only the case of frequency
degenerate OPO, i.e. when the frequency of the signal and idler are the
same; this also implies that $\gamma_x \simeq \gamma_y$ (without loss of generality
we set $\gamma_x=\gamma_y=1$ in our simulations, thus scaling time with the cavity
lifetime). The condition $a_x=a_y=a$ can be exactly obtained for example by
introducing compensating prisms in the cavity, as suggested in Appendix I.
Strictly speaking if $\omega=0$ the solutions (\ref{tw}) are not traveling waves
but rather standing phase waves; however, we will generally refer to them as TW.
It is important to note that considering the case $\omega=0$
is not restrictive in the sense that all features observed for this case
survive also for $\omega \neq 0$, at least when a uniform pump and
periodic boundary conditions are used. We checked this by integrating eqs.
(\ref{master1},\ref{master2}) also for $a_x \neq a_y$; the main difference is
that phase waves travel.
Under the hypothesis made in this paragraph the effective detuning simply reads
$\tilde{\Delta}= \Delta + a |{\bf k}|^2$.

TW solutions (\ref{tw}) exist for $|\mu|^2>1+\tilde{\Delta}^2=|\mu_c|^2$ if
$\sigma'+\sigma'' \tilde{\Delta}<0$; in this case only the solution
with the plus sign is acceptable in eq. (\ref{det-c}).  If
$\sigma'+\sigma'' \tilde{\Delta}>0$ solutions exist for
$|\mu|^2>|\mu_c|^2 - \frac{1}{|\sigma|^2} (\sigma'+\sigma''
\tilde{\Delta})$ and there is bistability up to $\mu_c$ among the
solutions obtained by taking the plus and minus signs. However,
when no mismatch is present ($\Delta k=0$), given that $\sigma'<0$, no
bistability regime exists and TW are found only for $|\mu|^2>|\mu_c|^2=1$.
Note that if no mismatch is present the solutions also satisfy $\alpha=0$
and therefore $A_x=A_y^*$. Hereafter we will consider only
the case of perfect phase-matching and leave the observations about
the effects of mismatch to section \ref{frem}.

A physical interpretation of the exact solutions can be given in terms of spatial
polarisation patterns. In fact, under the assumptions made, the Stokes parameters
\cite{phot}
\begin{eqnarray}
\label{def-stokes}
S_0 & = & |A_x|^2+|A_y|^2 \nonumber \\
S_1 & = & |A_x|^2-|A_y|^2 \nonumber \\
S_2 & = & A_x A_y^*+A_x^* A_y \nonumber \\
S_3 & = & i(A_x A_y^*-A_x^* A_y)
\end{eqnarray}
associated with the TW's at frequency $\omega_s=\omega_i=\omega_p/2$
can be simply calculated as
\begin{eqnarray}
\label{stokes}
S_0=2C, \;\;\;\; S_1=0, \;\;\;\; S_2=2 \; \;C \; cos(2{\bf k_c} \cdot
{\bf r}-\alpha), \;\;\;\; S_3=2 \; C \;sin(-2{\bf k_c}
\cdot {\bf r}+\alpha)
\end{eqnarray}

These equations mean that the coupled-TW solution $(A_x,A_y)$ represents an
elliptically polarized field with a two-valued spatially periodic azimuth
($\theta=tan^{-1}(S_2/S_1)/2$) and with a
spatially periodic ellipticity parameter $\eta({\bf r})=\alpha/2-{\bf
k_c} \cdot {\bf r}$.  In particular, in the direction of the
wave-vector ${\bf k_c}$ the polarisation changes from linear (with
$\theta=\pi/4$), to right circular, to linear (with $\theta=-\pi/4$),
then to left circular and back again to linear (with $\theta=\pi/4$)
with a spatial period $\lambda_c/2=\pi/k_c$ (see fig. \ref{polar}).

Other stable solutions can be found as combinations of the basic TW solutions
(\ref{tw}) as shown in ref. \cite{long96a} for a triply resonant, non-degenerate
OPO. They can be generally written as:

\begin{equation}
\label{solut}
\left( \begin{array}{c} A_x \\
A_y^* \end{array} \right)=(f_1 exp(i k_c x) +f_2 exp(i k_c y) +
f_3 exp(-i k_c x)+f_4 exp(-i k_c y))
\left( \begin{array}{c} 1 \\
1 \end{array} \right)
\end{equation}

The coefficients $f_k$ determine the kind of solution; for example the
TW is $(f_1,f_2,f_3,f_4)=(\sqrt{C},0,0,0)$ and the standing rolls are
$(\sqrt{C},0,\sqrt{C},0)$.  An interesting solution for this case is
that of the alternating rolls
$(f_1,f_2,f_3,f_4)=(\sqrt{C},i\sqrt{C},\sqrt{C},i\sqrt{C})$ i.e.:

\begin{equation}
\label{square}
\left( \begin{array}{c} A_x \\
A_y^* \end{array} \right)=2 \sqrt{C} (cos(k_c x)+i \; cos(k_c y))
\left( \begin{array}{c} 1 \\
1 \end{array} \right)
\end{equation}

that gives rise to a square intensity pattern.
We found that this structure is stable with periodic
boundary conditions, but it deforms and disappears as soon as other,
more realistic, boundary conditions are applied. We show in fig.
\ref{alt-rolls} the square array with periodic boundary conditions
after an evolution of $t=500$ time units; no appreciable change of the
initial condition (\ref{square}) has been observed.

The equations also have a trivial steady-state solution $A_x=A_y=0$, that
represents the non lasing state of the OPO and whose stability analysis is
similar to that performed in refs. \cite{oppo94,oppo94a,long96b}. Briefly,
$A_x=A_y=0$ becomes unstable for $|\mu|>1$ if $\Delta<0$ and for
$\mu>\sqrt{1+\Delta^2}$ if $\Delta>0$.
For $\Delta>0$ the trivial solution becomes unstable for homogeneous
perturbations ($k_c=0$) and no pattern or structure is expected to
be spontaneously formed. In numerical simulation we observed the
transition to a final, stable, homogeneous state of intensity $C$,
preceded by a transient regime where domain walls form but soon disappear
\cite{oppo98}. For $\Delta<0$ the most unstable modes are TW with a critical
wave-vector $|{\bf k_c}|^2=-\Delta/a$, i.e. such that the equivalent
detuning $\tilde{\Delta}$ is zero. Note that the most unstable modes are
also exact TW solutions of eqs. (\ref{master1},\ref{master2}).
Although stability of such TW's was not studied, numerical solutions give
evidence that they are stable and moreover that they are spontaneously
formed starting from a random perturbation of the trivial steady-state
when the pump amplitude is above the threshold of the parametric interaction.
Then, the spontaneously selected TW's have wave-vector modulus
$|{\bf k_c}|=k_c=\sqrt{-\Delta/a}$ but a random orientation or phase in different
region of space; this causes the occurrence of defects in the patterns as will
be shown below. Hereafter we will treat only the case of polarization
pattern formation, i.e. $\Delta<0$.

If the pump amplitude is weakly above the threshold ($5 \%$) two different final
situations can be found starting from a random perturbation of the trivial
steady-state under periodic boundary
conditions. In the first, all defects initially formed tend to
annihilate each other and the final state is a single TW of random orientation,
whose wave-vector is exactly $k_c$. As in the case of non-degenerate OPO's
\cite{long96b} no intensity patterns appear but only phase stripe patterns,
which correspond to the polarisation patterns described by eqs. (\ref{tw}) and
(\ref{stokes}). In another case, for the same pump amplitude but with other initial
random conditions, ordered structures of defects may form as shown in fig.
\ref{random_order}. By looking at the real and imaginary parts of the
field $A_x$ it is clear that these defects are found along the fronts
which separate regions where TW's with the same wave-vector but with
different phase have been selected. From the figure it is also clear
that these defects appear on a background of TW solutions, at those
points where the stripes do not match.

Defects also appear spontaneously with pump amplitudes well above threshold
($50 \%$); in this case the positions of the defects are not ordered, as
shown for example in fig. \ref{random_disorder} a) and b).
Defects always appear in both orthogonal polarisation components (i.e. the
signal and the idler simultaneously) and therefore they can be classified as
vectorial, since the two components of the vector field $(A_x,A_y)$ are zero at
the same point in space.
In fig. \ref{random_disorder} c) and d) the phase of the polarisation components
is also presented; note that it is not defined at the points where the
amplitude goes to zero. Then, such objects require a topological classification
that is furnished in the next section.
Finally note that setting $a_x \neq a_y$ or $\gamma_x \neq \gamma_y$
does not alter the basic features observed;
defects persist and are advected by the phase waves, which now travel as
previously described. This has been cheked by running the same simulation
(i.e. that of fig. \ref{random_order}) with and without equal coefficients.

It is always useful to have in mind the general characteristics of the
instability of the zero state in our OPO model, in order to relate it to
other pattern forming systems. In our case, the instability for $\Delta
< 0$ is at a non vanishing wavenumber $k_c$, so that a Swift-Hohenberg
equation for the complex envelope of the unstable modes is the
appropriate amplitude-equation description close to threshold. For the
case $a_x=a_y$, this Swift-Hohenberg equation has real coefficients.
This equation is derived from our basic equations (1,2)  in Appendix
II.

\section{Defect classification, formation and dynamics}
\label{defe}

In this section we analyse the formation and dynamics of defects, in
Type II OPO's. The existence of defects in a Type I non-degenerate
triply resonant OPO was pointed out in ref. \cite{valc97} where the
dynamics of an advected defect pair was also reported.
Here, we present a detailed study of the defects which
spontaneously appear in the Type II frequency degenerate OPO. The analysis
includes: a) the classification of the defect type, showing that these
defects are dislocations of the TW pattern; b) the formation process under
several different conditions and regimes and c) the trapping and
stabilization of a single defect.

A dislocation of a scalar complex field $A(r,\theta)$, $(r,\theta)$ being
spatial polar coordinates, can be defined as a configuration which can be
deformed continuously (in a neibourhood of a point ${\bf r_0}$, which is
then identified as the core of the dislocation and it is taken as the
origin of the polar coordinates) into the following function:

\begin{equation}
\label{dis-def}
A(r,\theta)=R(r) \; exp(i(\theta+k r \; cos \theta))=R(r) \; exp(i \phi)
\end{equation}

The function $R(r)$ is real and depends on the radial
coordinate such that $R(0)=0$.  This condition is necessary because the phase
$\phi$ of the field is not defined in the origin of coordinates. In fig.
\ref{disloc} a) we show the real part of $A$ from which we can observe
that the defect is found at the point where the stripes of the
background, whose wave-vector far from the defect is $k$, do not
match.  The phase $\phi$ is shown in fig. \ref{disloc} b); the defect is centered
at the point ${\bf r_0}=0$, around which the phase changes by $2 \pi$.
The topological charge associated with the phase change is defined
by the integral of the phase gradient on a closed contour surrounding
the point ${\bf r_0}$

\begin{equation}
n_0=\frac{1}{2 \pi} \oint_{{\bf r_0}} \nabla \phi \cdot d {\bf r} = 1
\end{equation}

Note that there exists also another dislocation obtained by setting
$\theta \rightarrow -\theta$ for which the topological charge $n_0$ is equal
to $-1$. However the most interesting feature of a dislocation is that
it is characterised by another topological charge. This is associated with
the function

\begin{equation}
\label{psifield}
\psi(x,y)=tan^{-1} \left( \frac{(\nabla \phi)_y}{(\nabla \phi)_x} \right)
\end{equation}

where the indices $x,y$ of the term on the right hand side refer to the vector
components. This function,
which defines the orientation angle of the wave-vector ${\bf k}$, is
shown in fig. \ref{disloc} c) for the field $A$ defined by eq.
(\ref{dis-def}). Note that there are two points where $\psi$ is not
defined: one is located at the point ${\bf r_0}$, corresponding to the
point of zero amplitude; the second is at another point ${\bf r_1}$
where $|{\bf r_1}-{\bf r_0}|=k^{-1}$. Then, the topological charges
associated with $\psi$ are:

\begin{equation}
m_0=\frac{1}{2 \pi} \oint_{{\bf r_0}} \nabla \psi \cdot d {\bf r} = 1
\end{equation}

where the integration contour surrounds the point ${\bf r_0}$ but leaves
outside the point ${\bf r_1}$ and

\begin{equation}
m_1=\frac{1}{2 \pi} \oint_{{\bf r_1}} \nabla \psi \cdot d {\bf r} = -1
\end{equation}

where the integration contour surrounds the point ${\bf r_1}$ but leaves
outside the point ${\bf r_0}$. Note that there is not a zero in the amplitude
$R$ at the point ${\bf r_1}$.

We now turn our attention back to the Type II OPO and compare the
defects which are found in the numerical solutions of eqs.
(\ref{master1},\ref{master2}) with the dislocation just defined.
In this section, to better compare numerical and theoretical results only
periodic boundary conditions and homogeneous pump beams are considered;
more realistic conditions will be explored in section (\ref{isol}) when
stabilization of a dislocation is considered.

The randomly generated defects of figures \ref{random_order} and
\ref{random_disorder} are dislocations on a background of TW's, although in the
second case it is difficult to recognize the background solutions since the
defects are dense in the integration window. However a fingerprint of TW is found
also for the data of fig. \ref{random_disorder} by observing the far field in
fig. \ref{farfield}. There, a ring of modes of radius $k_c$ is clearly observable,
a clear indication that the background is made mostly of TW solutions of different
orientation but the same wave-number $k_c$. The dislocations can be also clearly
identified when the quantities $\psi_x$ and $\psi_y$ (defined again through eq.
(\ref{psifield}) where the phases are those of $A_x, A_y$) are calculated for
the same data of fig. \ref{random_disorder}. The result is presented in figures
\ref{random_psi} a) and b) where a contour plot of $|A_x|,|A_y|$ taken
at a low intensity level to indicate the locations of the zeros of the field,
is superimposed on the grey-scale coded images of the functions $\psi_x$
and $\psi_y$. From fig. \ref{random_psi} we note that in this static condition
``chains" of defects of the functions $\psi_{x,y}$ are formed. Two zeros of the
amplitude (defects of charges $n_0,m_0$) are in fact connected by the relative
pair of defects of the functions $\psi_{x,y}$ at the second point (defects
of charge $m_1$). This stems from the fact that the phase difference between
$A_x$ and $A_y$ is zero (when there is no mismatch) and thus $A_x=A_y^*$.

Regarding the dynamics \cite{movies} of the
spontaneously generated defects, we observe that they appear at the
early stage of the process. Some of them are
annihilated within a few thousand cavity lifetime units.  After this stage we
cannot observe any annihilation, up to 100000 cavity lifetime units.  The
initial number of dislocations and their positions change with different
initial conditions.  The statistical distribution of the final number
of defects, by changing initial conditions, is shown in fig.
\ref{stat}; observe that a Gaussian fit is actually very good.
We note that the defect density never reaches that of the
square pattern (eq. (\ref{square})) with critical wave-vector $k_c$.
After the initial formation stage, we observe in our runs that defects
have a slow, random motion. This regime is mostly dominated by the
attraction of pairs of defects of opposite charges $n_0=\pm 1$. For
longer times the motion tends to stop and an equilibrium condition is reached.
Although no detailed statistical study has been conducted we also note, by
inspection of fig. \ref{random_psi}, that there seems to exist a
characteristic length of separation between the defects, which is an
indication that an equilibrium is likely to be reached.

To better study defect interaction we use (\ref{dis-def}) as an initial condition
and integrate eqs. (\ref{master1},\ref{master2}). Due to the periodic
boundary conditions a pair of defects is generated, as shown in fig.
\ref{pair} a). These correspond to two dislocations in the
real part of each field (fig. \ref{pair} b)), two phase singularities
with opposite topological charge (fig. \ref{pair} c)) and two pairs
of defects associated with the quantity $\psi_x$ (fig. \ref{pair}
d)). At the initial stage the zeros of the amplitude, which have opposite $n_0$
charge, attract each other; later, when the $\psi_x$ defects (which have the same
charge $m_0=1$) get closer, this attraction is stopped and an equilibrium
position is found \cite{movies}. This paired structure is stable at least up to
the time explored by numerical solutions.

In general, the detailed behavior of defects dynamics is governed both by local,
phase curvature effects and by defect-defect interactions. The range of
these interactions depends on the parameters of the system, which
determine the spatial extent of the defect.
For example if we make the detuning more negative (e.g. $\Delta_x=\Delta_y=-1$ and
$a_x=a_y=1$) two defects displaced parallel to the background wavefronts
either annihilate, if close enough, or repel each other, if further away.

Figure \ref{my_fig2} also shows a pair of defects (same parameter values as
in Figure \ref{pair} but different initial displacement) undergoing
motion; the
background has a wave-vector of $2 k_c$ and it is stable when the system is
pumped far above threshold. When their separation is small enough, the defects
briefly rotate around each other before continuing their transverse motion.
As a consequence of this motion and the periodic boundary conditions, the
defects collide and annihilate shortly afterwards (not shown).

Figure \ref{my_fig1} shows another interesting example. Two defects approach
each other while the background phase wave undergoes a zig-zag
instability \cite{RMP}. When they meet they form a bound pair which
follows the local curvature of the background phase fronts, changing
direction as the slow phase dynamics alters this curvature. The bound defect
pair is stable for the duration of the simulation (115000 cavity
lifetimes). It also exists for lower pump powers, down to at least five
percent above threshold. On the other hand, at slightly  higher pump
powers than that used in Figure \ref{my_fig1}, the structure
destabilises:  after a short time the defect pair rotates slightly
before annihilating.

The previous examples show that long--lived defect interactions are
important features of the dynamics of Type II OPO's. This fact is in
agreement with instances of long term survival of many defects generated
by random initial conditions (e.g. Fig. \ref{random_disorder}) and is
clearly different from vortex dynamics observed, for example, in
complex Ginzburg-Landau models \cite{Pismen}.

Close to threshold, it is possible to study the behavior of
the Type II OPO dislocations by means of amplitude equations.
These are derived in Appendix II where comparisons
are made with simulations of the full model.

\section{Trapping of isolated dislocations}
\label{isol}

We know that periodic boundary conditions always require
the total topological charge within the domain to be zero.
In this section, we take into account that the pump amplitude $E_0$ actually has
a spatial dependence and periodic boundary conditions are not applicable. If a
spatial Gaussian or super-Gaussian distribution is used for the parametric gain
factor $\mu$ (eq. (\ref{mu})), the spontaneously generated defects often tend to
move to the region where the field is zero. The trapping of single
defects necessary for their experimental observation as single entities
is then an interesting issue.

We tried, successfully, to isolate a single defect in a
more realistic pump beam, by using, as an initial condition for one of
the fields, a doughnut mode.  This idea follows the experimental
results of induced dislocation formation in quadratic nonlinear
interactions \cite{berz98}. Note that no cavity was used in
those experiments and the defect was maintained by forcing a doughnut
mode in the field at the crystal input. Here the dislocation is
"written" at the beginning but, after that, no injection is required
since the cavity provides the necessary feedback to
sustain the structure. The resulting, stable trapped defect is shown in
fig. \ref{stable}. The spiral wave in the phase field spins when $a_x \neq a_y$,
otherwise it is stationary after the initial formation stage.
Note that in order to trap the defect the beam size must be
kept smaller than the wavelength of the most unstable TW
$\lambda_c=2\pi/k_c$ otherwise more defects are induced or the initial
defect is dragged outside the beam.  The
creation of other defects is due to the tendency of the system to
generate pairs. If the system size is comparable with the critical wavelength,
the paired defect will appear inside the pumping region and will destabilize the
original defect. If the size is smaller the second defect will be generated
outside the pumping region and therefore will not influence the
dynamics of the trapped defect (fig \ref{stable} c)). The amplitude
profiles of the fields of this single defect are shown in fig.
\ref{profile}.

This way of generating the defects is also useful for understanding the
mechanism of their formation.  The presence of a defect in one field,
say $A_x$, induces a defect of opposite charge $n_0$ in the other because
of the parametric coupling. In fact if $A_x \simeq 0$ at $t=0$ the term
which dominates the dynamics of $A_y$ in the initial regime is the
largest linear one, i.e. $\mu A_x^*$. Hence, $A_y$ is forced by the
complex conjugate of $A_x$ and a defect of opposite charge is formed.
Eventually the other terms of eqs. (\ref{master1},\ref{master2}) become
significant: the real part of the cubic term provides the saturation
which stabilizes the solution and nonlinear phase modulations appear if
mismatch is present, due to its imaginary part.

If we initialise a defect with topological charge $n_0=\pm 2$ using
the technique described above, we observe that it breaks up into
two defects of charge $n_0=\pm 1$.

Since the generation of such defects has been seeded externally,
the question arises whether it is possible to obtain defects
for positive detunings. In spite of the absence of stable
traveling--wave solutions, single vortices can be trapped
by using the same seeding technique.

We also observe that walk-off removes all structures from the pumping region
and a single TW is selected asymptotically.
However, it is interesting to note that in the transient regime
dislocations are formed.  They appear at the front which divides two
regions where stripes with different wave-vector modulus are selected.
The difference in the wave-vectors is due to the walk-off, as shown in
ref. \cite{josab}.

Finally, we estimate the threshold for the observation of this isolated
dislocation in an OPO. Considering data for $KTiOPO_4$ \cite{tang95}:
$\chi=7.33 \, pm/V, n_0=1.8$ at $\lambda_0=1.064 \mu m$, and by using the definition
of the effective coupling paramater defined in Appendix I ($\alpha$) and considering
a nonlinear crystal length of $1cm$, a mirror transmittivity $10^{-2}$, the
input field at threshold (E=1) can be evaluated as $E_{IN,0} \simeq
2 \, 10^4 \, V/m$ that yields an intensity $I=n_0/2 \, (\mu_0/\epsilon_0)^{1/2}
|E_{IN,0}|^2 \simeq 1 MW/m^2$  ($\mu_0, \epsilon_0$ are respectively the
vacuum magnetic permeability and electric permittivity constants).
The super-Gaussian beam used in the simulation has a beam diameter of about
10 normalized units; a spatial normalized unit corresponds, for the diffraction
coefficient of the simulation to about $0.195 mm$; hence the CW threshold
power is about $0.95W$. The physical cavity decay rates $\gamma_{x,y}$ are about
$0.17 GHz$ while the cavity detuning is $34MHz$. All these data are compatible
with an experimental realization.

\section{Mismatch effects}
\label{frem}

This section is dedicated to the study of the effects of mismatch
($\Delta k \neq 0$) between the fields in the OPO. As
previously noted, in real devices it is often possible that the selection
rules of the oscillation frequencies force the OPO to emit radiation with a
slight mismatch \cite{ecka91}.

The effects of the mismatch, predicted by the analysis of section
\ref{equat}, are the following: a) an increase in the threshold for
signal generation (i.e. the instability, see eq. (\ref{mu})); b) a spatial shift
of the iso-polarisation lines for the exact TW solutions. The latter effect
is due to the appearance of the phase shift $\alpha \neq 0$ (see eq.
(\ref{tw})) between polarisation components;
this stems from the imaginary parts of the coefficients $\mu$ and $\sigma$, the
contribution of the latter being a nonlinear phase modulation due to
the $\chi^{(2)}:\chi^{(2)}$ cascading effect \cite{steg95}.

For the case of the spontaneous generation of defects we have been able
to determine that the phase shift $\alpha$ is non zero in the regions
where the background solution dominates. The numerically found value is
very close to the value predicted by eq. (\ref{det-psi}). However,
approaching the zeros of the amplitude the phase difference also tends
to zero although the phase of the fields is not strictly defined at the
defects. In practice the zeros of the spatial function $\alpha(x,y)$
are located exactly at the positions of the field defects.  This
behavior can be explained as follows: close to the defects the cubic
terms tend to zero; in particular the term proportional to the
imaginary part of $\sigma$ (i.e. the Kerr-like phase modulation term)
is the first that can be neglected. Although $\mu=|\mu| exp(i \beta)$
is complex, by an appropriate re-definition of the variables as
$\tilde{A_j}=A_j exp(i \beta/2)$ $j=1,2$ we can obtain a set of
equations for $\tilde{A_j}$ where the parametric gain is purely real
($\mu''=0$).  Thus, we reduce the problem to the case where no mismatch is
present (see eq. (\ref{det-psi})) and therefore $\alpha \rightarrow 0$.

The final number of spontaneously generated dislocations seems also to
be affected by the phase mismatch. We note that this number
decreases by changing the mismatch from negative to positive values,
keeping all the other parameters fixed. In particular it seems that
positive values of the mismatch encourage the annihilation of the
pairs of dislocations.  This has been verified by numerical integrations
with the same parameters and initial conditions as fig. \ref{pair}
but with non zero mismatch.  For $\Delta k>0$ we observed that the pair
is annihilated, while for $\Delta k<0$ the pair reaches a stable
configuration. A possible cause of this asymmetry is the phase modulation
$\alpha$ introduced by the mismatch, which changes sign according to
the sign of the mismatch (see fig. \ref{c-psi}).

The effect of phase mismatch on the trapped defect is to destabilise
it, by generating asymmetries in the fields for both positive and
negative values of $\Delta k$.

\section{Conclusions}
\label{concl}

In conclusion we have studied the polarisation pattern and vectorial defect
formation in a Type II frequency degenerate optical parametric oscillator.

We found that the preferred solutions, those that are selected out of
an initial perturbation of the zero state, are conjugate traveling
waves in the two components of the polarisation.  A spatial
polarisation pattern is then formed: the state of polarisation changes
along the spatial coordinate parallel to the selected wave-vector
${\bf k_c}$ with period equal to $\lambda_c/2=\pi/k_c$. In particular
the state of polarisation changes along the meridian of the Poincare
sphere which passes through the linearly polarized states of azimuth
$\theta=\pm \pi/4$ if $\Delta k=0$.  The magnitude of the wave-vector of
the selected solution is fixed by the parameters and has a single random
orientation close to threshold and multiple random orientations far from
threshold.
Combinations of traveling waves, that form square patterns are also
stable solutions and, close to the threshold of the instability,
ordered arrays of defects can form spontaneously.

Such defects are isolated zeros of the two linear components of
the polarisation, i.e. they are vectorial defects.  They
are dislocations which form in spatial positions where the background
solutions (traveling waves) do not match spatially.  Two different kinds
of topological charge must be defined: one kind of charge is associated
with the phase and two charges with the director angle of the field
wave-vector.  The first charge, located at the point where the
amplitude goes to zero, can be $\pm 1$ and is always opposite in the
two polarisation components.  The charge associated with the
wave-vector is always $+1$ at the point where the amplitude is zero and
$-1$ at the paired point, where the amplitude is not zero.  The
polarisation components are totally correlated; in fact all defects,
both in the phase and wave-vector fields, have a corresponding defect
in the other polarisation component.  In this way the defects form
chains in which the separation among defects seems to have a typical
size, which is of the order of the background wavelength.

The trapping of an isolated defect has also been
demonstrated. This is accomplished by keeping the size of the pump beam
smaller than the critical wavelength of the preferred traveling wave
such that a second defect cannot be created inside the
pump beam but rather lies outside and does not influence the dynamics.

Finally, we have addressed the effects which may arise when the nonlinear
interaction is slightly phase mismatched, i.e. the increase in the threshold of
signal generation, and the linear and nonlinear phase shifts among the
polarisations, the latter due to the cascading effect. Numerical results
showed that a positive mismatch can favor the annihilation of dislocations.

Possible applications of defects can be foreseen in the field of particle
and atom trapping \cite{swar96}.

\acknowledgements This work is supported by the European Commission through the Project
QSTRUCT (ERB FMRX-CT96-0077); Financial support is also acknowledged from
EPSRC (UK)(Grants Nos. GR/M 19727 and GR/M 31880), SHEFC (UK) (Grant VIDEOS) and
MCyT (Spain) (Project BMF2000-1108). G-LO acknowledges support from SGI.

\section{Appendix I}
\label{app1}
In this appendix we derive the dynamical equations for a cw Type-II
OPO in the presence of diffraction.
We start from the amplitude equations for the pump, signal and idler
fields in the crystal \cite{zhan95}
\begin{eqnarray}
\label{AEq}
\partial_z E_0 + \frac {n_0}{c} \,\,\, \partial_t E_0 &=&
\frac{i}{2k_0} \nabla^2 E_0 - \frac {4 \pi \Omega_{0} \chi} {n_0 c}
\,\,\, E_x E_y \,\,\, e^{-i \Delta k z} \nonumber \\
\partial_z E_x + \frac {n_x}{c} \,\,\, \partial_t E_x &=&
\frac{i}{2k_x} \nabla^2 E_x + \frac {4 \pi \Omega_{x} \chi} {n_x c}
\,\,\, E_0 E_y^* \,\,\, e^{i \Delta k z} \\
\partial_z E_y + \frac {n_y}{c} \,\,\, \partial_t E_y &=&
\frac{i}{2k_y} \nabla^2 E_y + \frac {4 \pi \Omega_{y} \chi} {n_y c}
\,\,\, E_0 E_x^* \,\,\, e^{i \Delta k z}. \nonumber
\end{eqnarray}
where $E_0, E_x, E_y$ are the slowly varying amplitudes of pump, signal
and idler respectively, $\Omega_0 = \Omega_x + \Omega_y$ is the
frequency constraint on the OPO, $k_i = n_i \Omega_i / c$ are the
wave-numbers, $\Delta k = k_0 - k_x - k_y$ is the
phase mismatch, $\chi$ is the second order susceptibility of the
crystal, and $c$ is the speed of light.
We treat here the perfectly matched case ($\Delta k = 0$) in order to
focus on the form of the diffraction coefficients of the final equations.
The phase matching condition implies that
once the three frequencies and two refractive indices are given, then
the third refractive index is determined. For example for $n_0$ we have
\begin{eqnarray}
n_0 = \frac {n_x \Omega_x + n_y \Omega_y}{\Omega_x + \Omega_y}
\;\;\;\;\;\;\;\;\;\;\;\;\;\;\;\;\;\;\;
n_0 = \frac {n_x + n_y}{2}
\end{eqnarray}
where the second equation is valid at frequency degeneracy.
It is also useful to introduce $\Omega_x = \mu \Omega_0$,
$\Omega_y = \nu \Omega_0$ with $\mu + \nu = 1$ and the effective
coupling parameter $\alpha = 4 \pi \Omega_0 \chi / (n c)$
to obtain
\begin{eqnarray}
\partial_z E_0 + \frac {n_0}{c} \,\,\, \partial_t E_0 &=&
\frac{i}{2k_0} \nabla^2 E_0 - \alpha  E_x E_y \nonumber \\
\partial_z E_x + \frac {n_x}{c} \,\,\, \partial_t E_x &=&
\frac{i}{2k_x} \nabla^2 E_x + \mu \alpha E_0 E_y^* \\
\partial_z E_y + \frac {n_y}{c} \,\,\, \partial_t E_y &=&
\frac{i}{2k_y} \nabla^2 E_y + \nu \alpha E_0 E_x^* . \nonumber
\end{eqnarray}

All these equations describing the fields in the crystal have a
similar form of the kind
\begin{eqnarray}
\label{PEQ}
\partial_z E_i + \frac {n_i}{c} \partial_t E_i = i d_i \nabla^2 E_i +
\beta_i \,\, NLT(E_j,E_m)
\end{eqnarray}
where $NLT$ stands for Non Linear Terms and the indices $i, j, m$ take
the symbolic values $0, x, y$.

We consider a ring cavity of legth ${\cal L}$ with a crystal length
$L$ and apply longitudinal boundary conditions and the standard Mean Field
Limit (MFL). We consider here the triply resonant case, the doubly-resonant
case of non resonated pump being a subset of this general frame. We then
obtain
\begin{eqnarray}
\partial_{t'} E_i + \frac {cL} {{\cal L} + (n_i - 1) L}
\,\,\, \partial_{z} F_i =
- \gamma_i F_i - i {\hat \delta_i} E_i + i a_i \nabla^2 E_i +
\gamma_i E_{IN,i} + {\hat \beta_i} NLT(E_j,E_m)\,
\end{eqnarray}
where
\begin{eqnarray}
\label{NPAR}
t'&=&t+\left[\frac{{\cal L}-L}{c}\right]\;\frac{z}{L} \nonumber \\
\gamma_i &=& \frac {c \; \sigma_i \; \varepsilon} {{\cal L} + (n_i - 1) L};
\;\;\;\;\;\;\;\;\;\;\;\; {\hat \delta_i} = \frac {c \; \delta_i}
{{\cal L} + (n_i - 1) L} = \frac {(\omega_i - \Omega_i) {\cal L}}
{{\cal L} + (n_i - 1) L} ; \nonumber \\
a_i &=& \frac {c {\cal L}} {2 k_i [{\cal L} + (n_i - 1) L]};
\;\;\;\;\;\;\;\;\;\;\;\;
{\hat \beta_i} = \frac {c \beta_i L} {{\cal L} + (n_i - 1) L}.
\;\;\;\;\;\;\;\;\;\;\;\;
\end{eqnarray}
Note that the presence of birefringence introduces an explicit dependence
on the refractive indices in the coefficients of the final equations. This is
a consequence of the tight comb of resonances observed when the
length of the cavity is scanned \cite{DSGF}.

Finally, we introduce a linear transformation of the fields
\begin{eqnarray}
A_0 = \frac{\alpha L}{\varepsilon} \sqrt{\frac {\mu \nu}
{\sigma_x \sigma_y}} F_0 \;\;\;\;\;\;\;\;\;\;\;\;
A_x = \frac{\alpha L}{\varepsilon} \sqrt{\frac {\nu}
{\sigma_0 \sigma_y}} F_x \;\;\;\;\;\;\;\;\;\;\;\;
A_y = \frac{\alpha L}{\varepsilon} \sqrt{\frac {\mu}
{\sigma_0 \sigma_x}} F_y ,
\end{eqnarray}
and a final normalisation of the parameters
\begin{eqnarray}
E(x,y) = \frac{\alpha L}{\varepsilon} \sqrt{\frac {\mu \nu}
{\sigma_x \sigma_y}} E_{IN,0}
\;\;\;\;\;\;\;\;\;\;\;\;
\Delta_i = \frac {\hat \delta_i}{\gamma_i} = \frac {\omega_i - \Omega_i}
{\gamma_i n_i} .
\end{eqnarray}
The final equations read
\begin{eqnarray}
\partial_{t'} A_0 &=& \gamma_0 \; [ - (1 + i \Delta_0) A_0 + E(x,y)
- A_x A_y] + i a_0 \nabla^2 A_0  \nonumber \\
\partial_{t'} A_x &=& \gamma_x \; [ - (1 + i \Delta_x) A_x
+ A_0 A_y^*] + i a_x \nabla^2 A_x \\
\partial_{t'} A_y &=& \gamma_y \; [ - (1 + i \Delta_y) A_y
+ A_0 A_x^*] + i a_y \nabla^2 A_y  \nonumber
\end{eqnarray}
where
\begin{eqnarray}
\gamma_i = \frac {c \; \sigma_i \; \varepsilon} {{\cal L} + (n_i - 1) L};
\;\;\;\;\;\;\;\;\;\;\;\;
a_i = \frac {c {\cal L}} {2 k_i [{\cal L} + (n_i - 1) L]} \, .
\end{eqnarray}
These equations clearly show that the loss and diffraction coefficients
depend critically on the refractive indices even in the frequency
degenerate case.

Since our analysis starts from stationary homogeneous solutions, the choice
of the temporal reference frame is fixed by the condition
$\Delta_x = \Delta_y=\Delta$ which excludes phase rotations for the
stationary homogeneous states \cite{DSGF}. In this case the final equations are
\begin{eqnarray}
\label{finalequ}
\partial_{t'} A_0 &=& \gamma_0 \; [ - (1 + i \Delta_0) A_0 + E(x,y)
- A_x A_y] + i a_0 \nabla^2 A_0  \nonumber \\
\partial_{t'} A_x &=& \gamma_x \; [ - (1 + i \Delta) A_x
+ A_0 A_y^*] + i a_x \nabla^2 A_x \\
\partial_{t'} A_y &=& \gamma_y \; [ - (1 + i \Delta) A_y
+ A_0 A_x^*] + i a_y \nabla^2 A_y  \nonumber \, .
\end{eqnarray}

We note that it is still possible to have equal loss and diffraction
coefficients for the three waves if we consider cavities
of different lengths ${\cal L}_i$ for each field. This can be
achieved by inserting compensating prisms of
chosen length $\Lambda_i$ and refractive indices $N_i$. In this case
the equations are the same as (\ref{finalequ}) but with redefined
loss and diffraction coefficients
\begin{eqnarray}
\gamma_i = \frac {c \; \sigma_i \; \varepsilon}
{{\cal L}_i + (n_i - 1) L + (N_i-1) \, \Lambda_i};
\;\;\;\;\;\;\;\;
a_i = \frac {c {\cal L}} {2 k_i
[{\cal L}_i + (n_i - 1) L + (N_i-1) \, \Lambda_i]} \, .
\end{eqnarray}
By adjusting losses $\sigma_i$ and compensating prism coefficients,
one can select equal $\gamma_i$. In this way the
compensating prism parameters can be left free to adjust the
diffraction coefficients $a_i$. In particular, we want $a_x = a_y$ at
degeneracy which means
\begin{eqnarray}
\label{cond}
n_x [{\cal L}_x + (n_x - 1) L + (N_x-1) \, \Lambda_x] =
n_y [{\cal L}_y + (n_y - 1) L + (N_y-1) \, \Lambda_y]
\end{eqnarray}
which leaves a lot of flexibility for the final setting.

Finally, we observe that in the absence of a cavity for the pump field
there are just two equations for the signal and idler waves remaining.
By setting $A_0=E-A_x A_y$ in equations (\ref{finalequ}), we obtain
equations (\ref{master1},\ref{master2}) for the oppositely polarized
fields $A_x$ and $A_y$ in a doubly-resonant configuration.

\section{Appendix II}
\label{app2}

In this Appendix we perform a weakly nonlinear analysis in the
region $k_c \ll 1$ and close to threshold in order to determine the
appropriate amplitude-equation description for the instability. Related
derivations have been presented in \cite{long96c,valc96}. For simplicity,
consider the case
\begin{eqnarray}
\gamma_x & = & \gamma_y = \gamma \nonumber \\
a_x & = & a_y = a
\end{eqnarray}
With the benefit of some {\em a posteriori} knowledge, we rewrite the
field equations in terms of the variables $F_1= A_x+A_y^*$ and
$F_2= A_x-A_y^*$:
\begin{eqnarray}
\partial_t F_1 & = & \gamma \left[ (E-1) F_1 -\frac{1}{4} F_1^*
\left( F_1^2 - F_2^2 \right) \right] +
i\left( a \nabla^2 - \gamma \Delta \right) F_2 \nonumber \\
\partial_t F_2 & = & \gamma \left[ -(E+1) F_2 +\frac{1}{4} F_2^*
\left( F_1^2 - F_2^2 \right) \right] +
i\left( a \nabla^2 - \gamma \Delta \right) F_1.
\end{eqnarray}
We define a smallness parameter $\epsilon$ and perform appropriate
scalings and expansions:
\begin{equation}
E = 1+\epsilon^2 P \; , \;\;\;\;\; T=\epsilon^2 t \; , \;\;\;\;\;
X= \sqrt{\epsilon} \;x \; , \;\;\;\;\; Y= \sqrt{\epsilon} \; y
\; , \;\;\;\;\; \Delta = \epsilon \delta \; , \;\;\;\;\;
F_i= \epsilon F_{i1} + \epsilon^2 F_{i2} + ... \label{SH_assump}
\end{equation}
We can then proceed order by order in $\epsilon$ in a straightforward
manner. At third order we have
\begin{eqnarray}
\partial_t F_1 & = & \gamma \left[ (E-1) F_1 -\frac{1}{4} |F_1|^2 F_1 \right]
-\frac{\gamma}{2} \left( \frac{a}{\gamma} \nabla^2 - \Delta \right)^2 F_1
\label{SHE} \\
F_2 & = & \frac{i}{2} \left( \frac{a}{\gamma} \nabla^2 -\Delta \right) F_1
\end{eqnarray}
The important point is that there is a single order parameter ($F_1$)
governed by a Swift-Hohenberg equation (SHE) \cite{RMP} and
a second field ($F_2$) which is slaved to $F_1$. In other words, the dynamics
of the system is described by a single complex scalar field. Note that it is only
when $A_x=A_y^*$ that the order parameter becomes real \cite{valc97}.
An extra imaginary transverse Laplacian term appears in (52) when $a_x
\neq a_y$.

\begin{figure}
\centerline{\psfig{figure=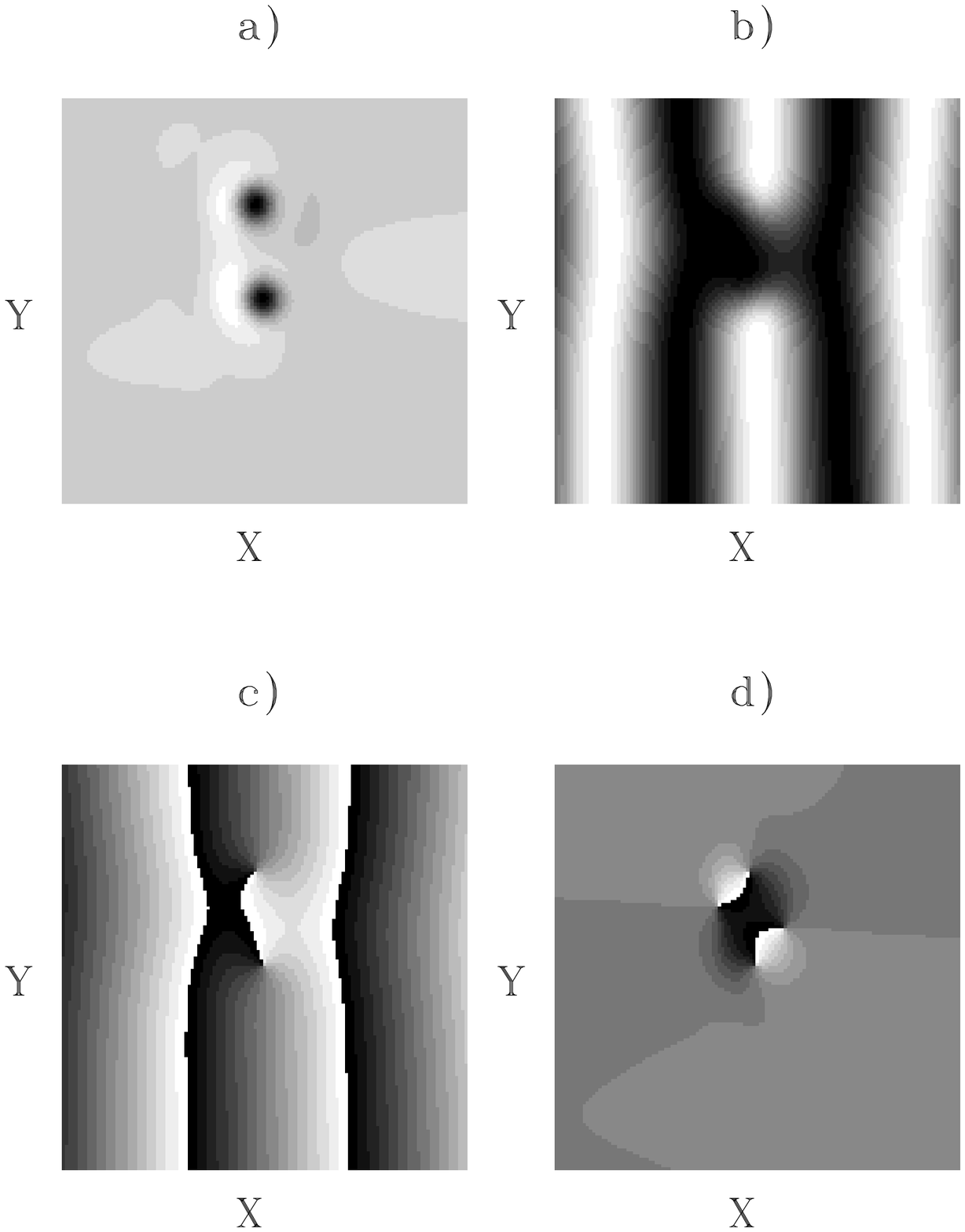}}
\vspace{9cm}
\caption{}
\label{pair}
\end{figure}

\begin{figure}
\centerline{\psfig{figure=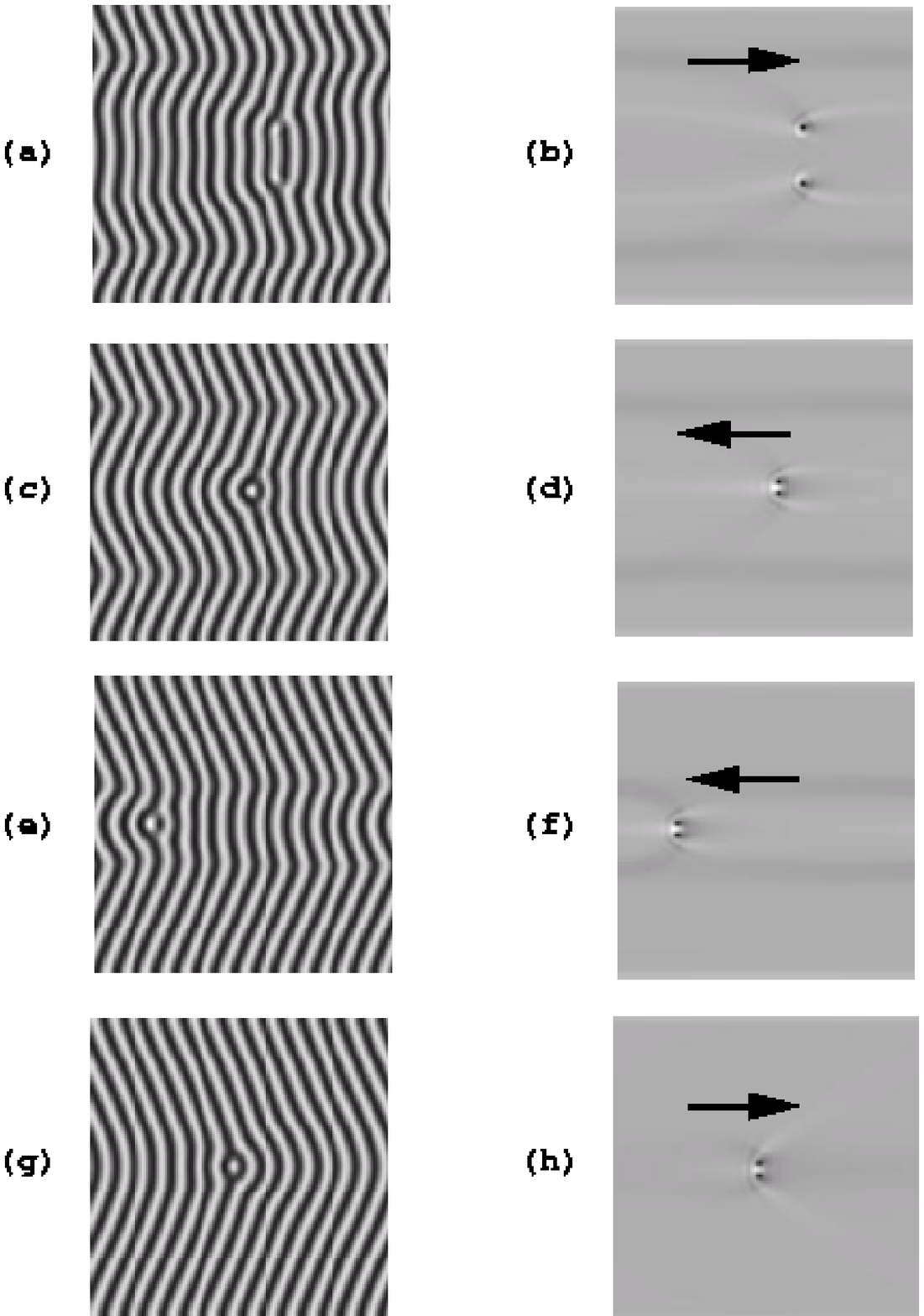}}
\caption{}
\label{my_fig1}
\end{figure}

Despite the assumptions underlying its derivation (Equations
(\ref{SH_assump})), the SHE can sometimes offer a good description of the system
even outside its expected range of validity. For example when integrating eq.
(\ref{SHE}) with the same parameter values as in Figure \ref{pair}, the final
state is essentially the same stable defect pair.
Nonetheless, notable failures of the SHE can be found; for example in reproducing
the movement of the bound defect pair in Figure \ref{my_fig1}.
This indicates that the original dynamics is not well reproduced when the
assumptions leading to (\ref{SH_assump},\ref{SHE}) are violated, and serves as a
warning about the limitations of order parameter equations such as (\ref{SHE}).

\newpage

\begin{figure}
\centerline{\psfig{figure=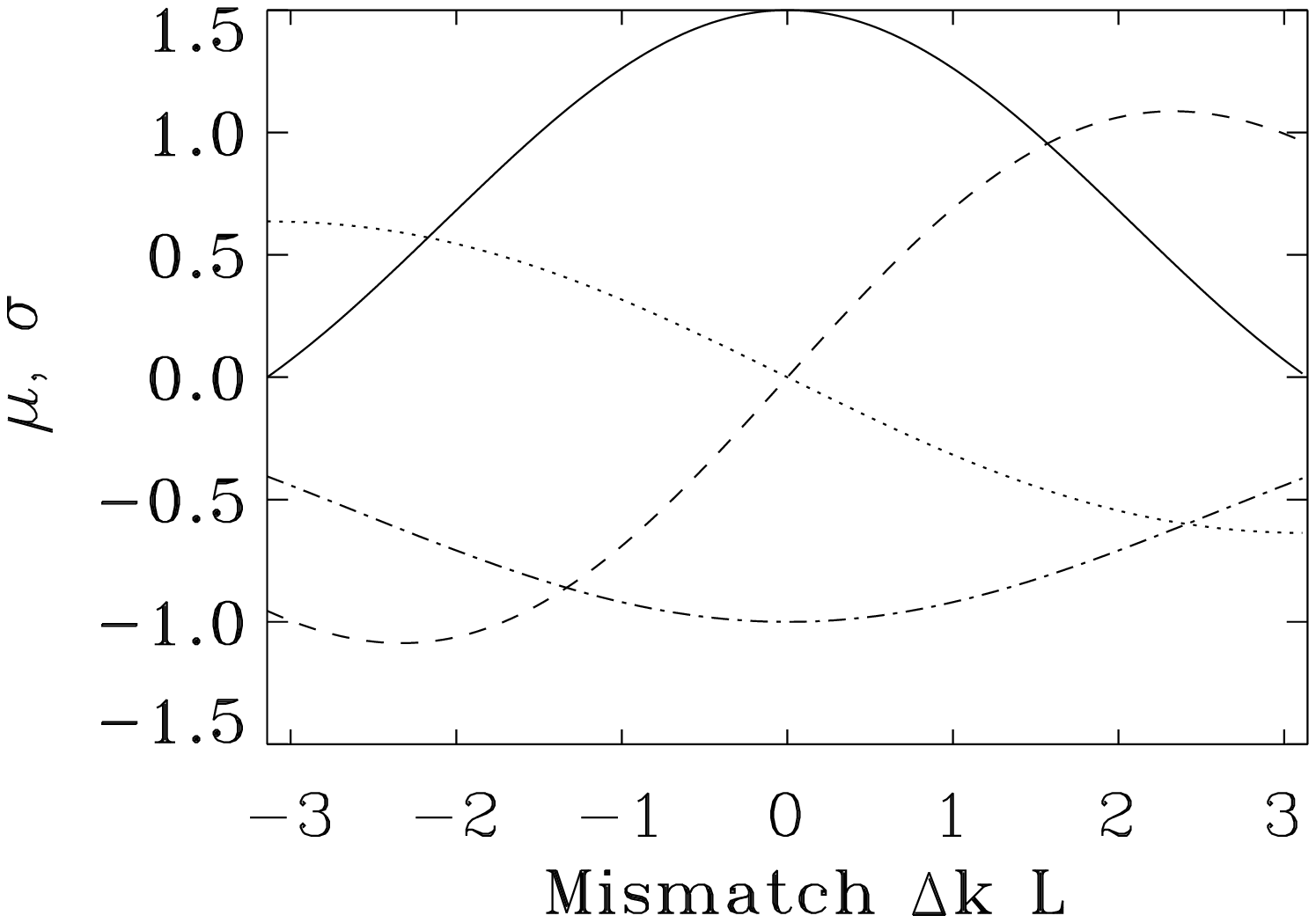}}
\caption{}
\label{reimcoef}
\end{figure}

FIGURE CAPTIONS:

Fig. \ref{reimcoef}: The solid (dashed) curve is the real (imaginary) part of
$\mu$ and the dashed-dotted (dotted) curve the real (imaginary)
part of $\sigma$ as functions of the mismatch ($E_0=1.5$).

\begin{figure}
\centerline{\psfig{figure=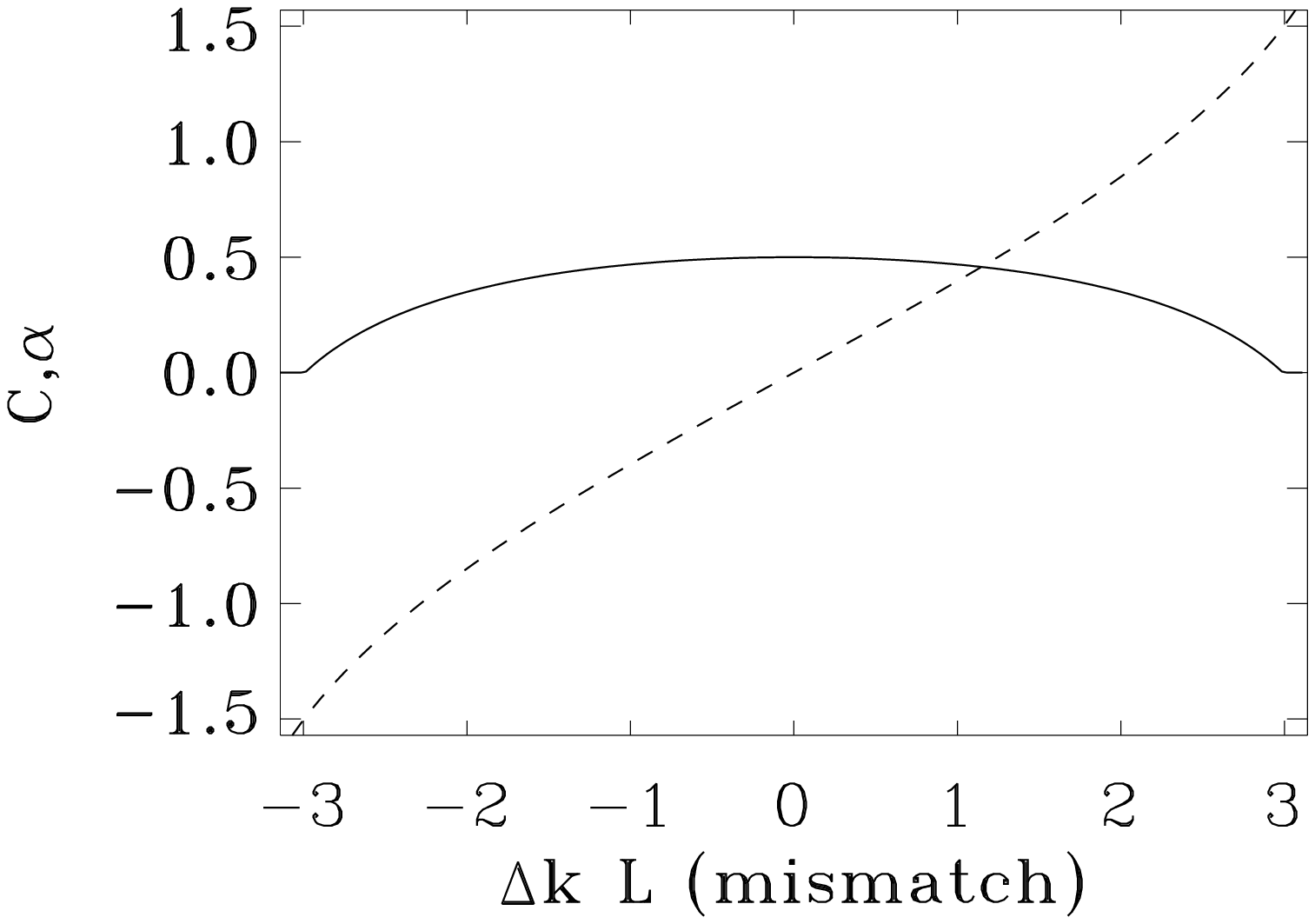}}
\caption{}
\label{c-psi}
\end{figure}

Fig. \ref{c-psi}: The solid (dashed) curve is the traveling wave solution
intensity $C$ (phase difference $\alpha$) as a function of the mismatch, as given
by eqs. (\ref{det-c},\ref{det-psi}) for $\tilde{\Delta}=0$.

\begin{figure}
\centerline{\psfig{figure=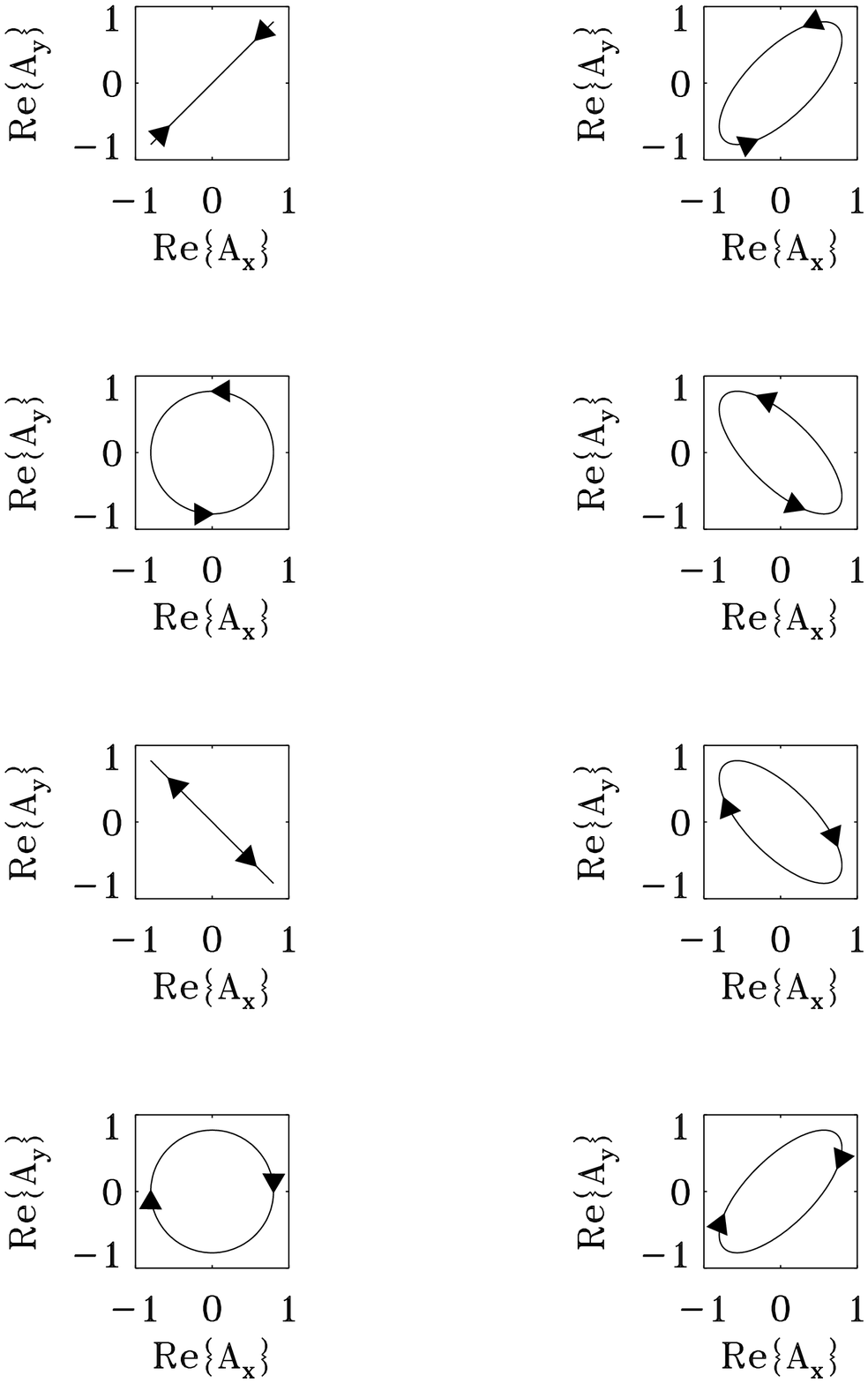,width=15cm}}
\caption{}
\label{polar}
\end{figure}

Fig. \ref{polar}: The change of the state of polarisation of the solution
(\ref{tw}) (${\bf k_c}=(k_c,0), C=0.8, \alpha=0, \phi_0=0$) is shown at the
spatial points of coordinates $x=n \pi/8, n=0..7$ (left to right and top to
bottom).

\begin{figure}
\centerline{\psfig{figure=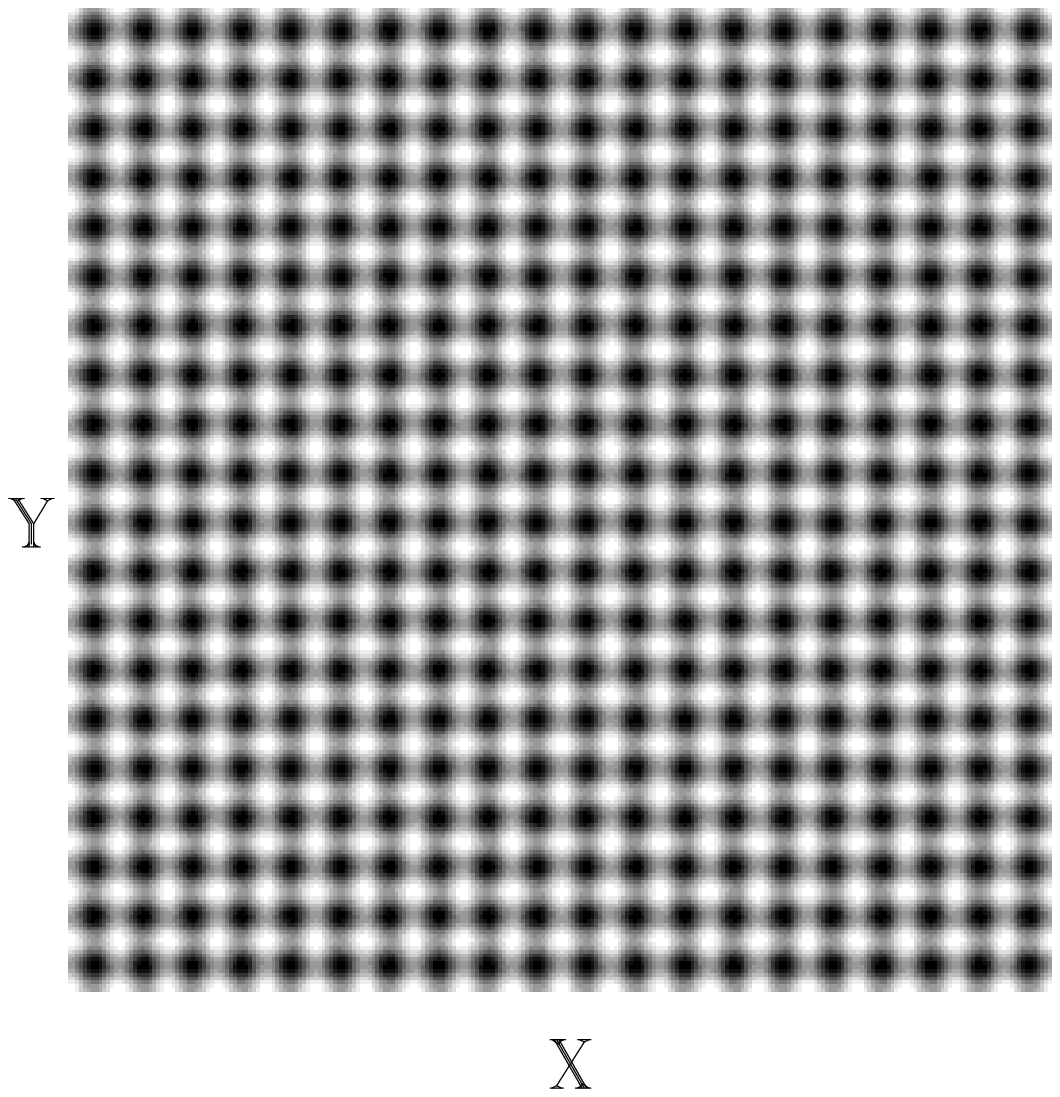}}
\vspace{3cm}
\caption{}
\label{alt-rolls}
\end{figure}

Fig. \ref{alt-rolls}: The amplitude ($|A_x|$) of the alternating roll solution
(\ref{square}) at $t=500$ time units ($a=0.8, \Delta=-0.8,
E_0=1.5$, the integrating window size was $X_L=Y_L=62.8$ units).

\begin{figure}
\centerline{\psfig{figure=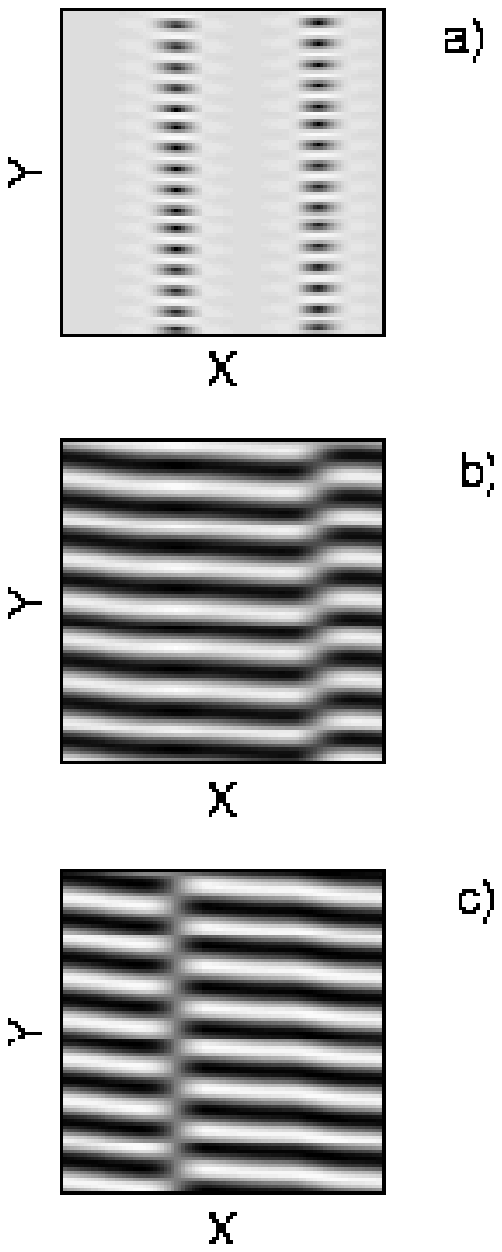,width=8cm}}
\caption{}
\label{random_order}
\end{figure}

Fig. \ref{random_order}: Ordered arrays of dislocations obtained close to
threshold ($E_0=1.05$): a) $|A_x|$; b) $Re(A_x)$; c) $Im(A_x)$. The
other parameters were $a_x=2, a_y=2.05, \gamma_x=1, \gamma_y=1.025,
 \Delta=-0.8, X_L=Y_L=80, t=4000$.

\begin{figure}
\centerline{\psfig{figure=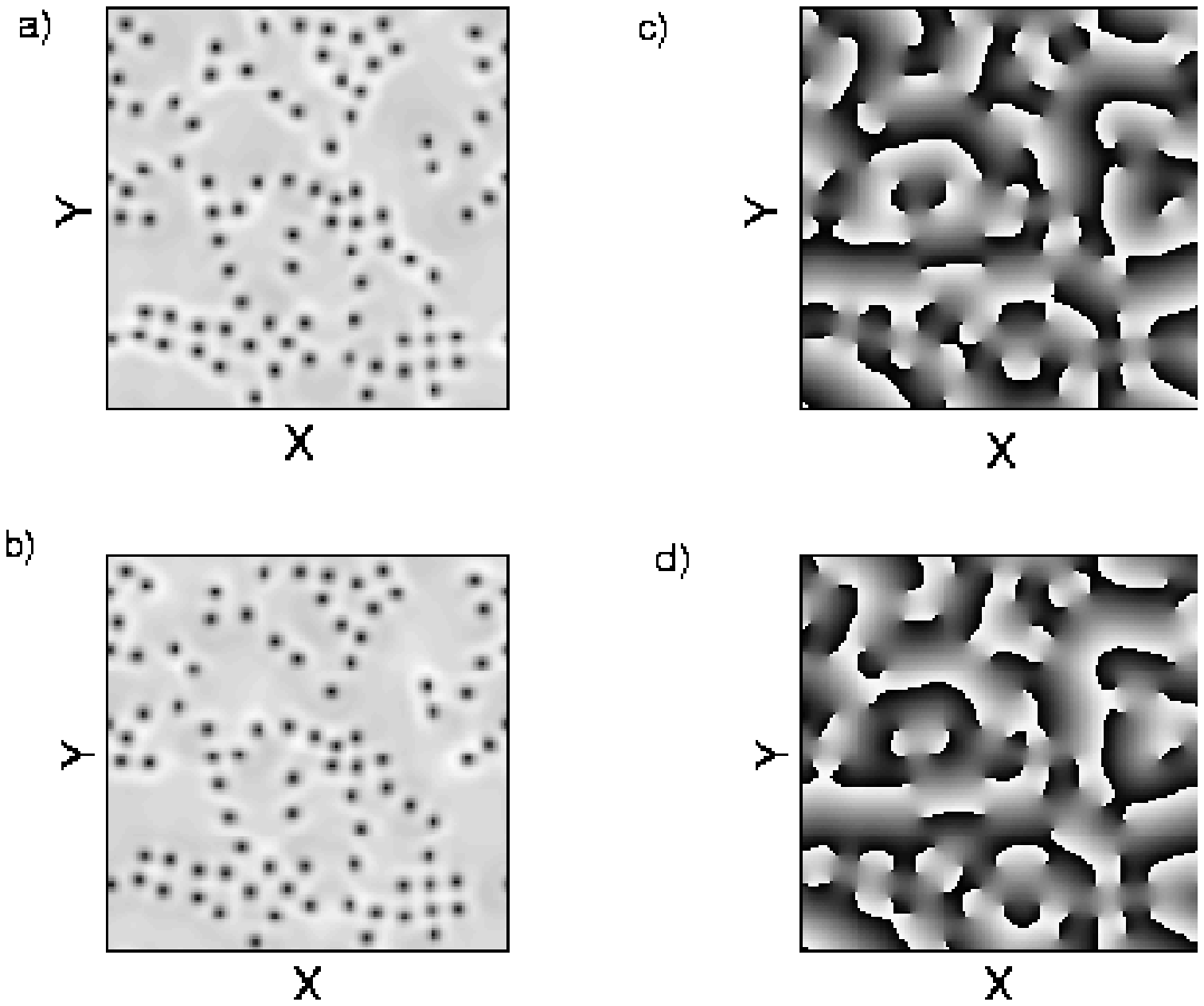}}
\caption{}
\label{random_disorder}
\end{figure}

Fig. \ref{random_disorder}: Amplitude and phase of the polarisation components
of the field well above threshold ($E_0=1.5, a=1, \Delta=-0.25, X_L=Y_L=80,
t=2000$). a) $|A_x|$; b) $|A_y|$: c) $tan^{-1}[Im(A_x)/Re(A_x)]$; d)
$tan^{-1}[Im(A_y)/Re(A_y)]$. Note that the defects are always
in pairs and that signs of the charge $n_0$ are opposite in the two components.

\begin{figure}
\centerline{\psfig{figure=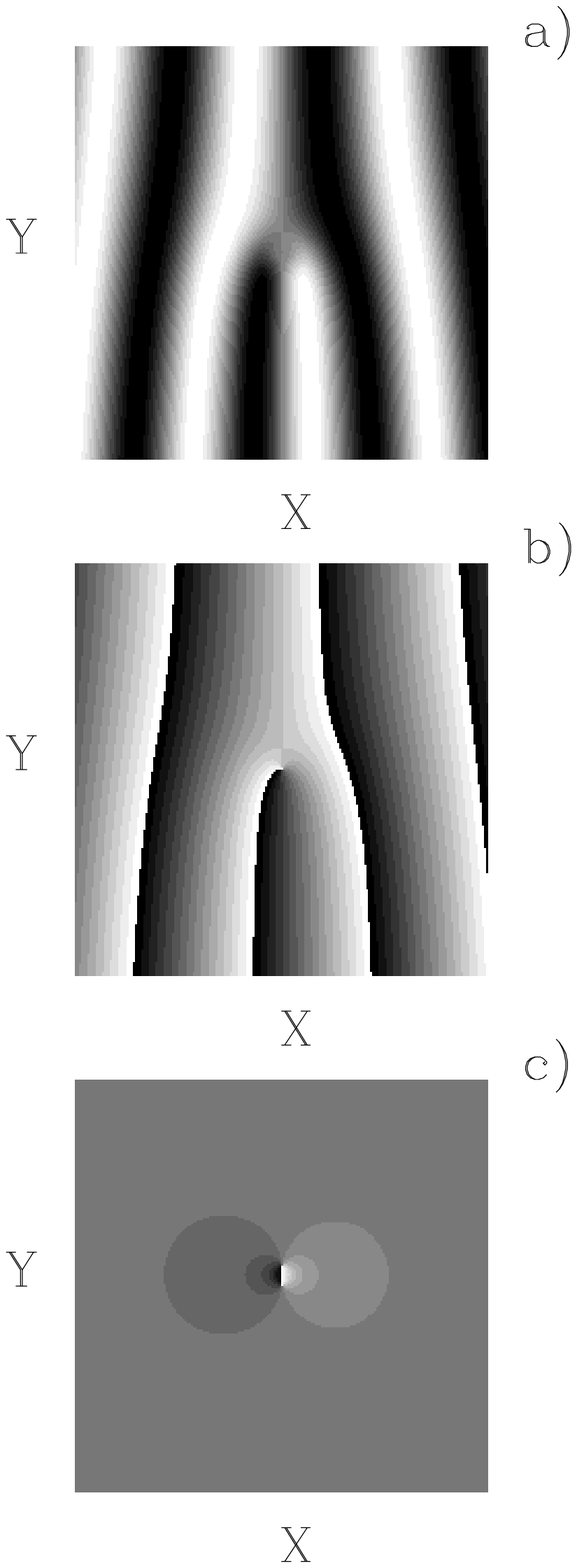}}
\vspace{9cm}
\caption{}
\label{disloc}
\end{figure}

Fig. \ref{disloc}: Dislocation of a complex field: a) $Re(A)$; b) $\phi$;
c) $\psi$; where $A,\phi$ are defined by eq. (\ref{dis-def}) and $\psi$ by
eq. (\ref{psifield}) ($k=1, R(0)=0, R(\infty)=1$).

\begin{figure}
\centerline{\psfig{figure=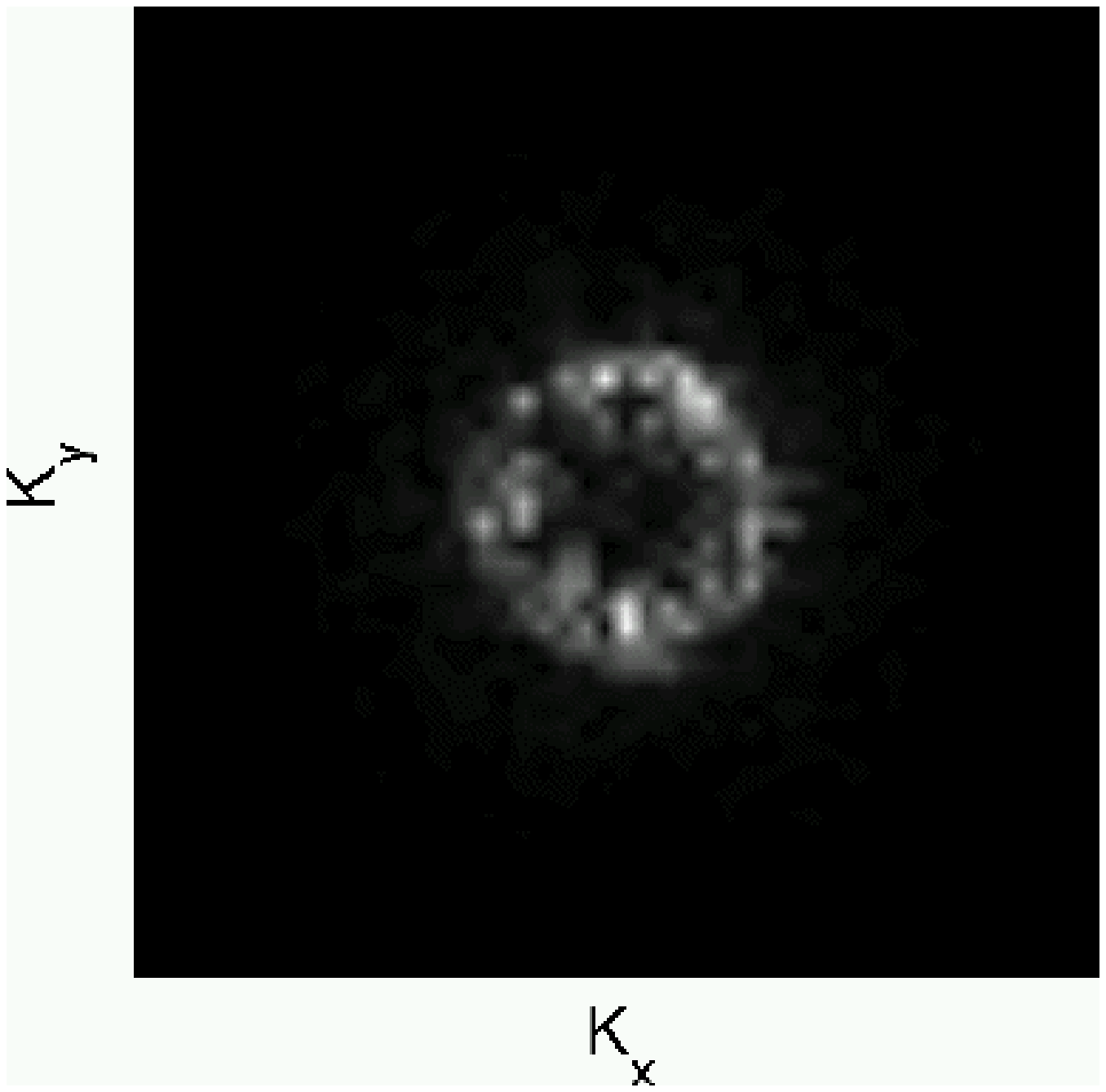}}
\vspace{3cm}
\caption{}
\label{farfield}
\end{figure}

Fig. \ref{farfield}: The amplitude of the far field (Fourier transform) of
$A_x$ for the same data of fig. \ref{random_disorder}.
The average modulus of the wavenumber of the modes in the ring is about 0.47
and the predicted value is $k_c=\sqrt{-\Delta/a}=0.5$.

\begin{figure}
\centerline{\psfig{figure=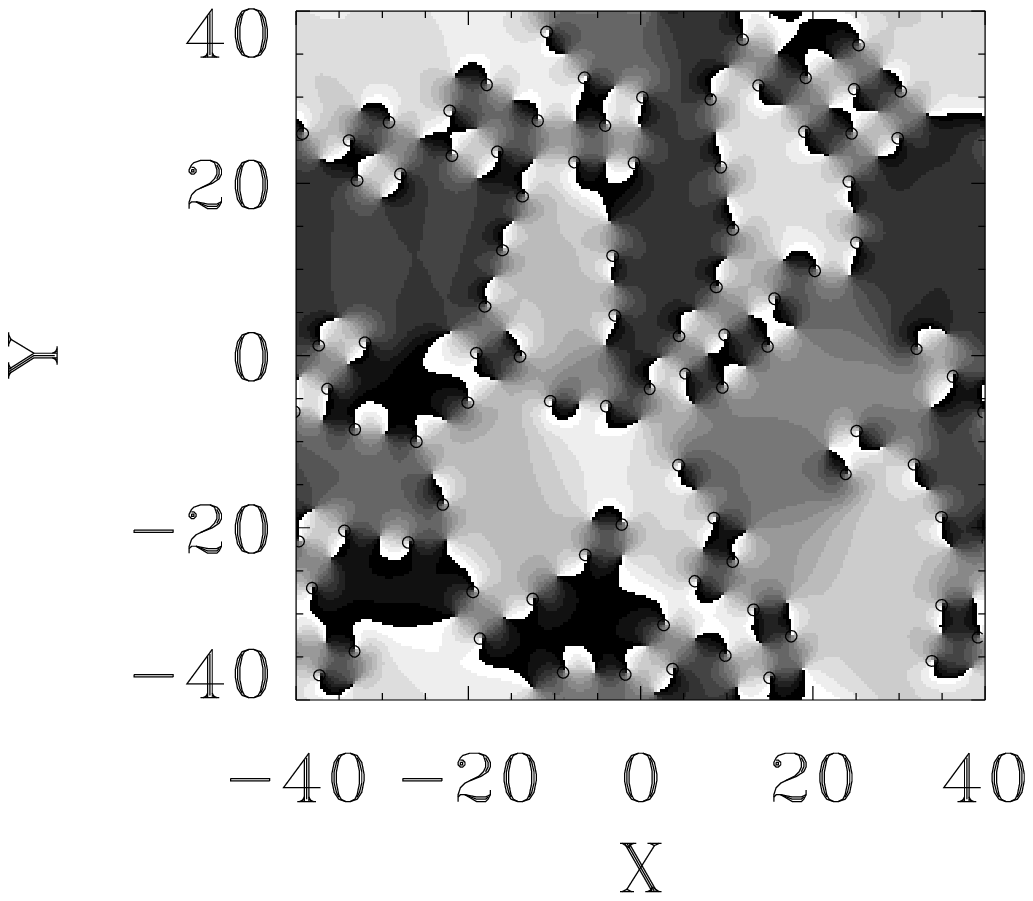,width=14cm}a)}
\vspace{-2cm}
\centerline{\psfig{figure=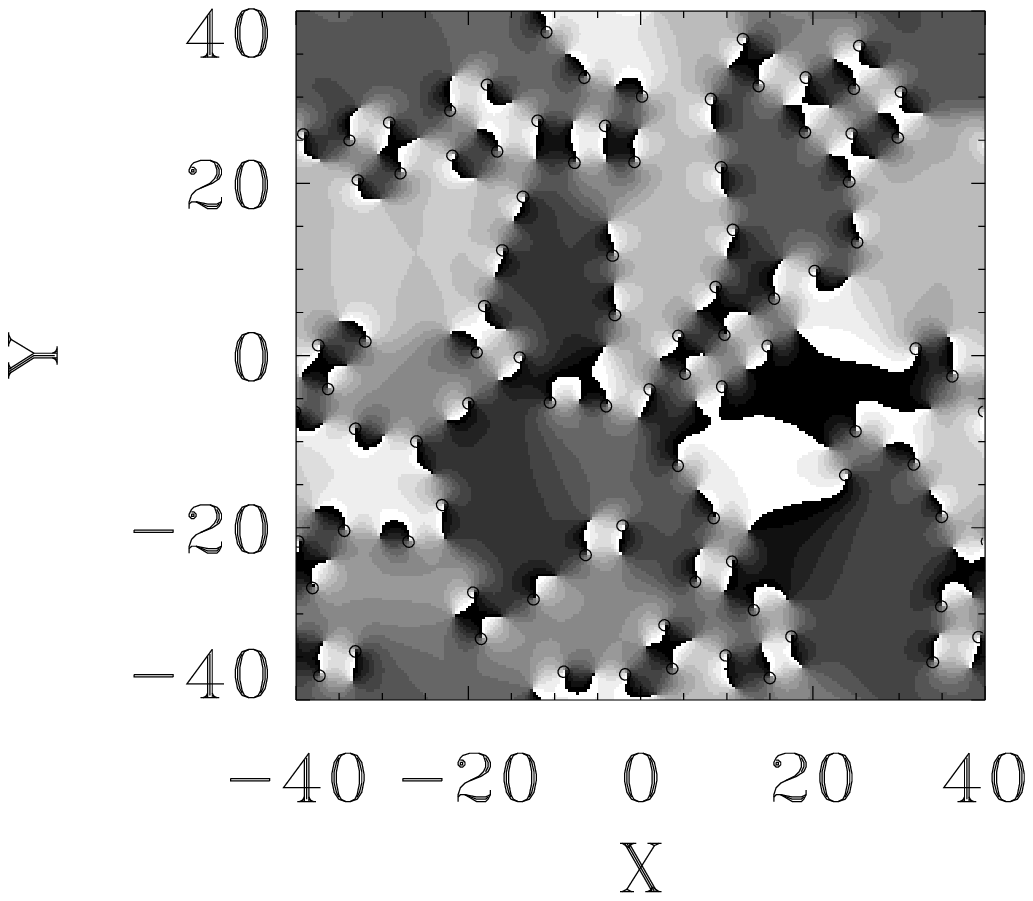,width=14cm}b)}
\vspace{3cm}
\caption{}
\label{random_psi}
\end{figure}

Fig. \ref{random_psi}: Functions a) $\psi_x$ and b) $\psi_y$ for the data of
fig. \ref{random_disorder}. Contour plots, with a contour level close to zero,
of $|A_x|,|A_y|$ have been superimposed to locate the positions of the amplitude
zeros.

\begin{figure}
\centerline{\psfig{figure=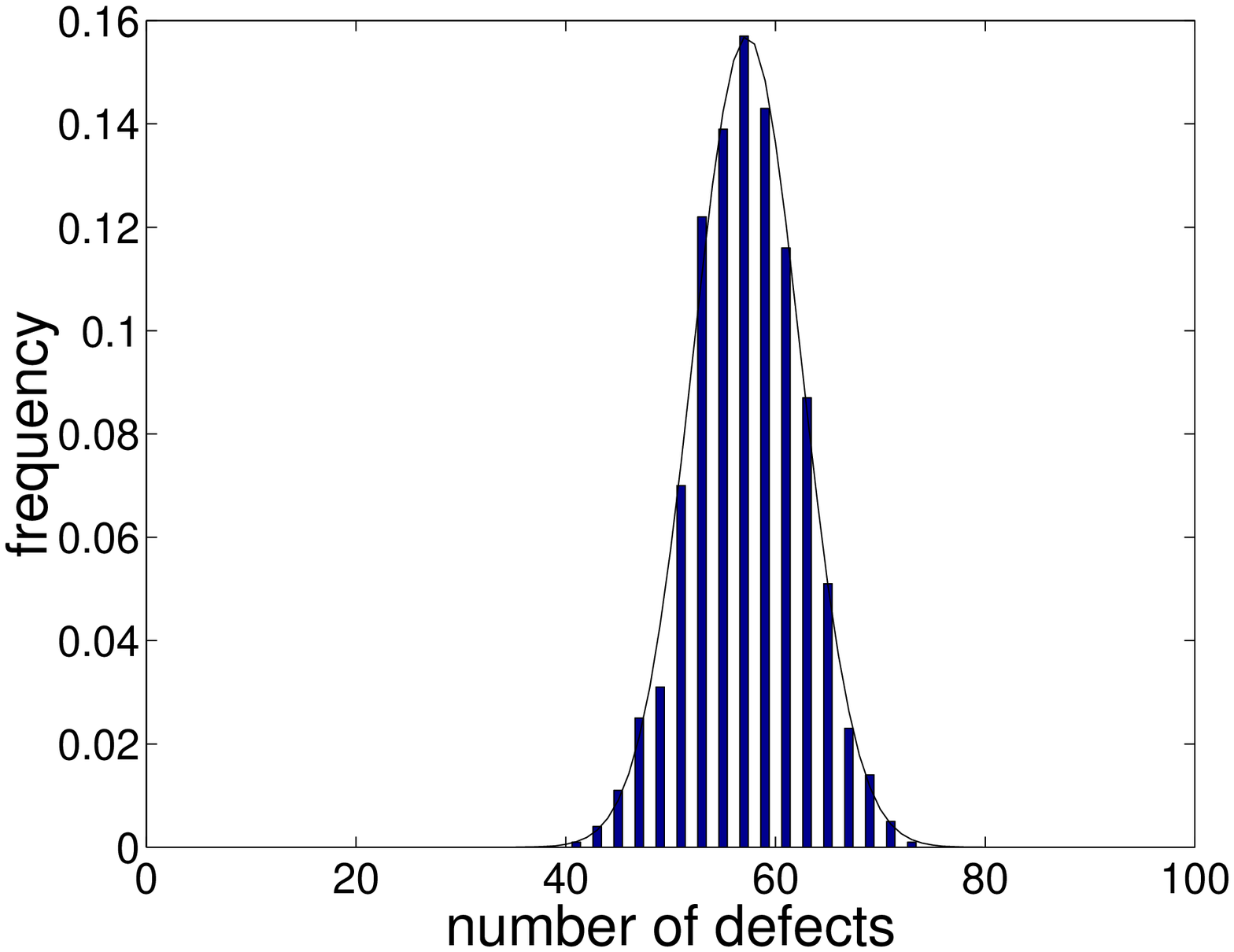}}
\caption{}
\label{stat}
\end{figure}

Fig. \ref{stat}: Distribution (bars) of the final number of defects for different
random initial conditions ($a=1, \Delta=-0.4, E_0=1.5, X_L=Y_L=50$). The
Gaussian fit average and variance are those obtained form the data.

Fig. \ref{pair}: Pair of dislocations obtained by integrating
(\ref{master1},\ref{master2}) with a single dislocation (\ref{dis-def}) as an
initial condition for $A_x$ and a random initial condition for $A_y$:
a) $|A_x|$; b) $Re(A_x)$; c) $\phi_x$; d) $\psi_x$. The integration
parameters were: $a=2, \Delta=-0.2, E_0=1.5, X_L=Y_L=100, t=2000$

\begin{figure}
\centerline{\psfig{figure=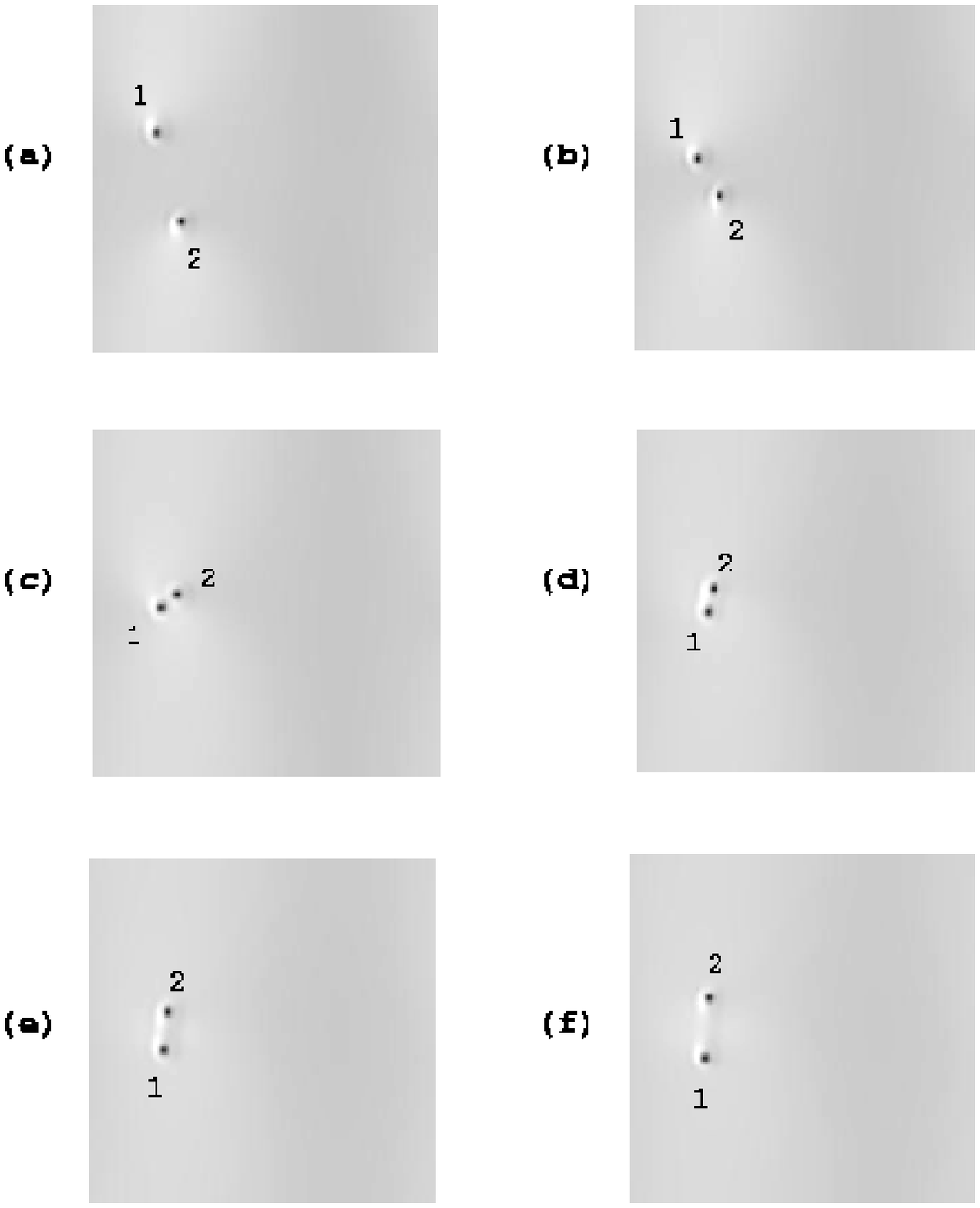}}
\caption{}
\label{my_fig2}
\end{figure}

Fig. \ref{my_fig2}: Interaction of two vortices (labelled $1$ and $2$)
for $\gamma_x=\gamma_y=1$, $\Delta=-0.2$ and $a_x=a_y=2$, $X_L=Y_L=16\pi/k_c$.
The images show $|A_1|$ at intervals of $50$ cavity lifetimes.

Fig. \ref{my_fig1}: A bound defect pair for $\gamma_x=\gamma_y=1$,
$\Delta=-1$ and $a_x=a_y=1$, $X_L=Y_L=32\pi/k_c$. Figures (a), (c), (e) and (g)
show the real part of $A_x$ at t=5500, 11000, 36000 and 54000 cavity
lifetimes respectively. The figures in the rightmost column show
$|A_x|$ at the corresponding times. The arrows indicate the direction
of transverse motion of the defects.

\begin{figure}
\centerline{\psfig{figure=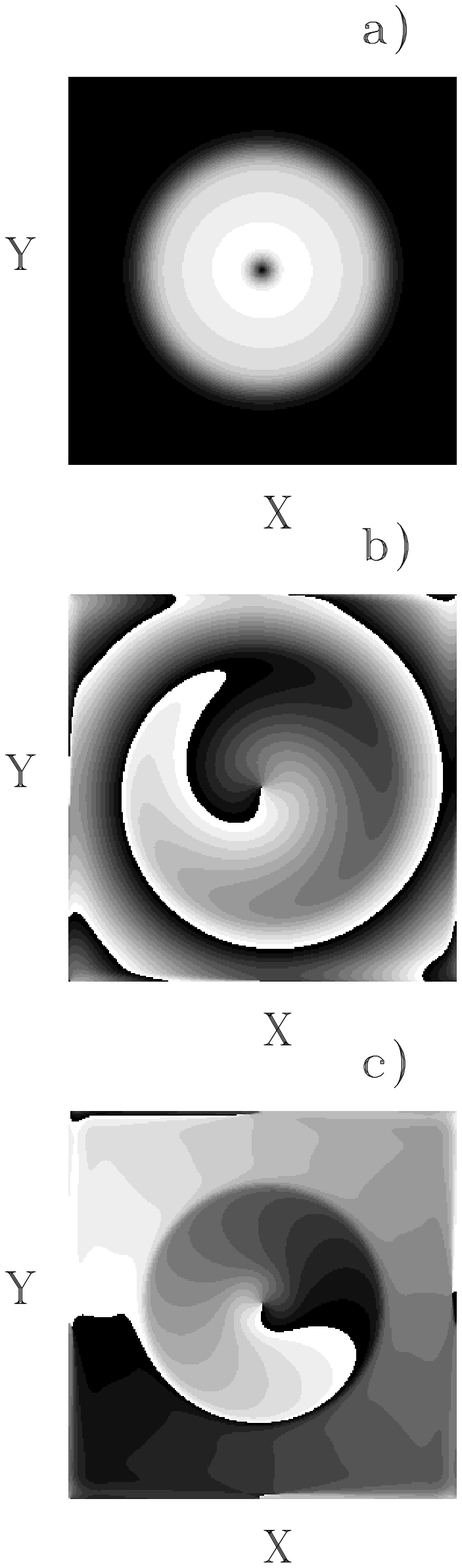}}
\vspace{9cm}
\caption{}
\label{stable}
\end{figure}

Fig. \ref{stable}: Isolated defect: a) $|A_x|$; b) $\phi_x$; c) $\psi_x$.
The initial condition for $A_x$ was a doughnut mode of the form
$A_x(x,y)=0.01 \; (x+iy) \; exp(-(x^2+y^2)/200)$ while $A_y$ was random.
The parameters of the integration were: $a=2, \Delta=-0.2, E_0=1.5, X_L=Y_L=60,
t=2000$.

\begin{figure}
\centerline{\psfig{figure=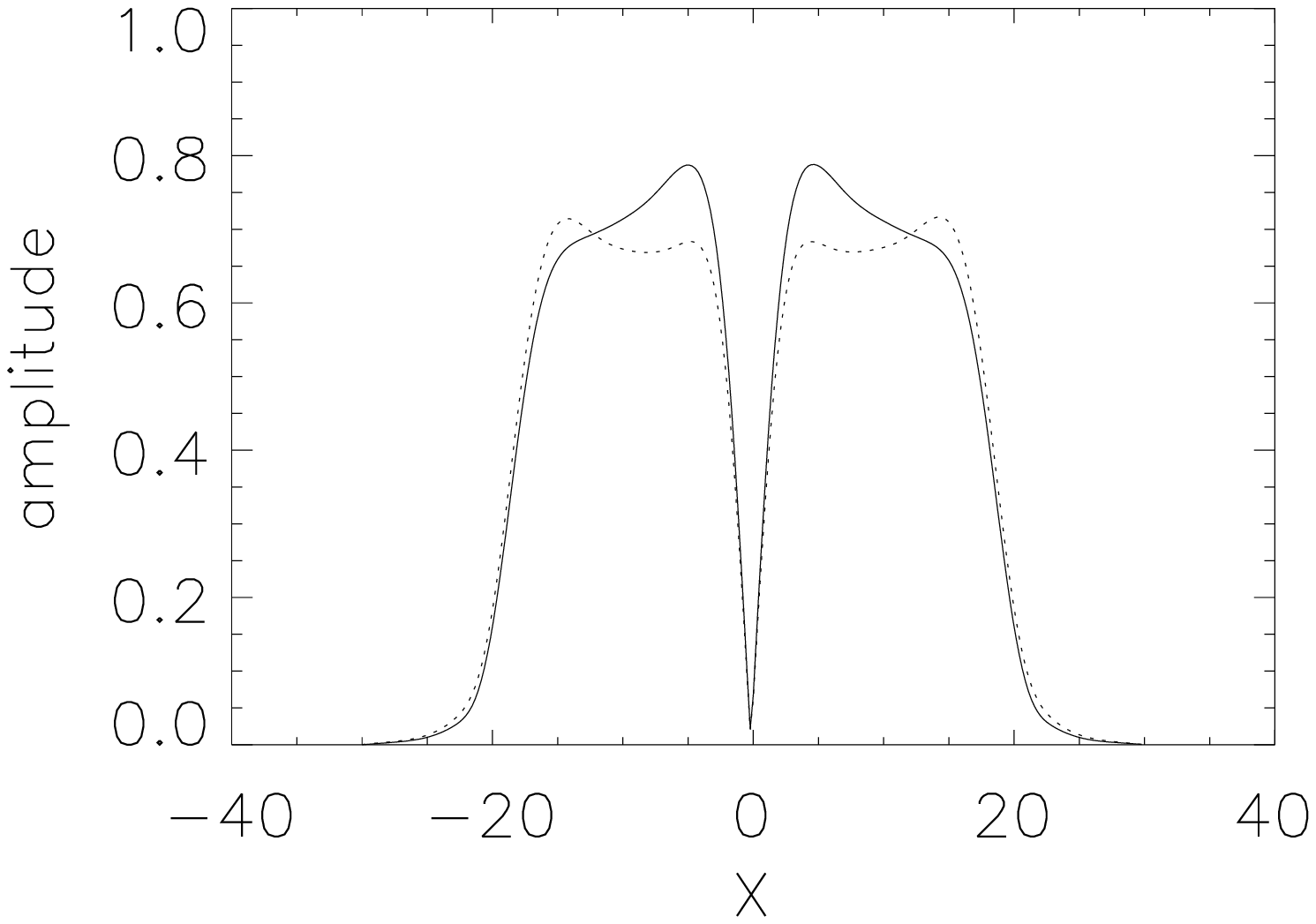}}\caption{}
\label{profile}
\end{figure}

Fig. \ref{profile}: Amplitude profile of the isolated defect of fig.
\ref{stable}: $|A_x|$ solid curve, $|A_y|$ dashed curve.

\newpage

\newpage

\newpage

\newpage

\newpage

\newpage

\newpage

\newpage

\newpage

\newpage

\newpage

\newpage

\newpage

\newpage

\newpage

\begin{thebibliography}{99}
\bibitem{note} In our approximation the cavity is almost filled by the
nonlinear crystal and thus the cavity length is very close to the crystal
length.
\end{thebibliography}
\end{document}